\documentclass[fleqn,usenatbib]{mnras}

\usepackage{newtxtext,newtxmath}

\usepackage[T1]{fontenc}

\DeclareRobustCommand{\VAN}[3]{#2}
\let\VANthebibliography\thebibliography
\def\thebibliography{\DeclareRobustCommand{\VAN}[3]{##3}\VANthebibliography}

\usepackage{graphicx}	
\usepackage{amsmath}	
\usepackage{amssymb}	

\defcitealias{roth+2023}{\textsc{CONGRuENTS} I}

\newcommand{\aref}[1]{\hyperref[#1]{Appendix~\ref{#1}}}

\usepackage{orcidlink}

\title[CONGRuENTS II]{\textsc{CONGRuENTS} (COsmic-ray, Neutrino, Gamma-ray and Radio Non-Thermal Spectra). II. Population-level correlations between galactic infrared, radio, and $\gamma$-ray emission}

\author[M. A. Roth et al.]{
Matt A. Roth,$^{\orcidlink{0000-0002-4204-5026}}$$^{1}$\thanks{E-mail: matt.roth@anu.edu.au (MAR)}
Mark R. Krumholz$^{\orcidlink{0000-0003-3893-854X}}$$^{1,2}$,
Roland M. Crocker$^{\orcidlink{0000-0002-2036-2426}}$$^{1}$,
and Todd A. Thompson$^{\orcidlink{0000-0003-2377-9574}}$$^{3}$
\\
$^{1}$Research School of Astronomy \& Astrophysics, The Australian National University, Canberra, Australian Capital Territory, 2611, Australia\\
$^{2}$ARC Centre of Excellence for All-Sky Astrophysics in
Three Dimensions (ASTRO-3D), Canberra, Australian Capital Territory, 2611, Australia\\
$^{3}$Department of Astronomy, Center for Cosmology \& Astro-Particle Physics, and Department of Physics, The Ohio State University, Columbus, OH 43210, USA
}

\date{Accepted 2024 March 25. Received 2024 March 17; in original form 2023 October 9}

\pubyear{2024}

\begin{document}
\label{firstpage}
\pagerange{\pageref{firstpage}--\pageref{lastpage}}
\maketitle

\begin{abstract}
Galaxies obey a number of empirical correlations between their 
radio, $\gamma$-ray, and infrared emission, 
but
the physical origins of these correlations remain uncertain. Here we use the \textsc{CONGRuENTS} model for broadband non-thermal emission from star-forming galaxies, which self-consistently calculates energy-dependent transport and non-thermal emission from cosmic ray hadrons and leptons, to predict radio and $\gamma$-ray emission for a synthetic galaxy population with properties drawn from a large deep-field survey. 
We show that our synthetic galaxies reproduce observed relations such as the FIR-radio correlation, the FIR-$\gamma$ correlation, and the distribution of radio spectral indices, and we use the model to explain the
physical
origins of these relations. Our results show that the FIR-radio correlation arises because the amount of cosmic ray electron power ultimately radiated as synchrotron emission varies only weakly with galaxy star formation rate as a result of the constraints imposed on gas properties by hydrostatic balance and turbulent dynamo action; the same physics dictates the extent of proton calorimetry in different galaxies, and thus sets the FIR-$\gamma$-ray correlation. We further show that galactic radio spectral indices result primarily from competition between thermal free-free emission and energy-dependent loss of cosmic ray 
electrons
to bremsstrahlung and escape into galactic halos, with shaping of the spectrum by inverse Compton, synchrotron, and ionisation processes 
typically
playing a 
sub-dominant
role. In addition to explaining existing observations, we use our analysis to predict a heretofore unseen correlation between the curvature of galaxies' radio spectra and their pion-driven $\gamma$-ray emission, a prediction that will be testable with upcoming facilities.
\end{abstract}




\section{Introduction}\label{sec:introduction}
Star-forming galaxies are an important source of broadband non-thermal emission.
Such emission is powered by radiative energy losses from the cosmic rays (CRs) that pervade their interstellar media (ISM). This emission is however outshone by thermal emission across much of the electromagnetic spectrum, tending to limit observational investigation of the non-thermal component to two windows, one in the radio and one at $\gamma$-ray energies. Here, non-thermal emission is significantly brighter and contamination by thermal processes either absent or significantly reduced. 

However, for simple physical reasons, one expects emission in these windows to correlate, at least roughly, with thermal emission, since both are ultimately powered mainly by recently-formed massive stars. These stars dominate thermal emission via their direct optical/UV and dust-reprocessed infrared emission, and they power non-thermal emission because most cosmic rays have their origin in ISM  shocks, 
injected by core collapse supernovae,
that accelerate particles up to ultra-relativistic energies. These ``primary'' CRs are made up of protons,
heavier ions 
(which we largely ignore for simplicity given their sub-dominance in shaping non-thermal emission),
and electrons.
We shall adopt here that 10\% \citep{1995STIN...9622970W, 2013A&A...553A..34D} of core collapse supernova energy gets deposited in the primary protons with another 2\% \citep{2010ApJ...717....1L} deposited in the primary electrons.
Accelerated protons go on to collide with the ISM, producing approximately equal numbers of $\pi^{0}$, $\pi^{+}$ and $\pi^{-}$ particles. The charged pions decay in a chain that eventually produces 
a considerable number of neutrinos, as well as
relativistic electrons and positrons -- ``secondary electrons" -- that add to the primary CR electron budget. Neutral pions, on the other hand, decay to two $\gamma$-ray photons, accounting for much of the $\gamma$-ray emission of a star-forming galaxy, with contributions from inverse Compton and bremsstrahlung emission from CR 
electrons\footnote{For simplicity, here and henceforth we refer to primary CR electrons and secondary CR electrons and positrons just as CR electrons, unless the distinction is important.} accounting for the balance. The same CR electrons also radiate significant synchrotron emission at radio wavelengths.

The expectation that galactic thermal and non-thermal emission should correlate is borne out -- surprisingly well -- by observations, most notably in the far infrared-radio correlation (FRC) and the far infrared-$\gamma$-ray correlation (F$\gamma$C). The FRC is a near constant, close to linear power-law, spanning nearly four decades in star-formation rate, that relates galaxies' far-infrared luminosity to their radio fluxes. This relation has been known observationally for several decades \citep{1971A&A....15..110V, 1973A&A....29..263V, 1985A&A...147L...6D, 1985ApJ...298L...7H, Condon92a, 2001ApJ...554..803Y}. When parameterised in the form of $S_{\rm radio} \propto L_{\rm FIR}^{k}$, early measurements of the index $k$ ranged from slightly sub-linear, $k \sim 0.94$ \citep{1985A&A...147L...6D}, to close to linear, $k \sim 1$ \citep{Condon92a, 2001ApJ...554..803Y}, to slightly super-linear, $k \sim 1.1$ \citep{1985ApJ...298L...7H}. More recent results indicate a consensus in favour of a slightly super-linear FRC, with \citet{2003ApJ...586..794B} finding $k \approx 1.10$ and \citet{2017ApJ...847..136B} finding $k \approx 1.15$[\footnote{This index results from converting their stated index of 1.27 for the correlation between radio luminosity and star formation rate to a radio-FIR correlation using the relation we obtain below for the FIR/star-formation rate relation.}].

While a rough FRC is to be expected, the observed tightness of the FRC is quite surprising given the large dynamic range in galaxy properties it covers, extending from dwarfs with star formation rates $\ll 1$ M$_\odot$ yr$^{-1}$ and magnetic field strengths of a few $\mu$G to the most intense starbursts that form stars at $\gtrsim 100$ M$_\odot$ yr$^{-1}$ and have magnetic fields in the mG range, implying a $\sim 6$ dex range in magnetic energy density. The range of star formation rates and gas masses per unit area present in our sample 
population of star-forming galaxies in CANDELS
similarly imply a $\sim 6$ dex span in radiation energy density, which directly affects the fraction of energy lost to inverse Compton emission, and a dynamic range of $\sim 4$ orders of magnitude in ISM density, which controls bremsstrahlung and ionisation losses. One might naively expect galaxies with such a disparate range of properties to differ significantly in the efficiency with which they reprocess CR energy to the radio, leading to large deviations or dispersion in the FRC, but such is not 
observed.

While the FRC has been known for decades, the related F$\gamma$C is significantly more difficult to observe, given the dearth of instruments capable of measuring the low $\gamma$-ray fluxes produced by star-forming galaxies. Observational results to date are based on a handful ($\sim 10$) of star-forming galaxies (excluding those that are known to house active galactic nuclei), mostly measured with the Fermi Gamma-ray Space Telescope. These galaxies are either very nearby, such as the Magellanic Clouds and M31, or moderately nearby and exceptionally luminous, such as starburst galaxies like Arp 220 and M82. The limited data set gathered thus far is often fit by a power law $L_{\gamma} \propto L_{\rm FIR}^{k}$, with indices reported in the literature indicating slightly superlinear relations, $k \approx 1.21$ \citep{2020A&A...641A.147K}, $k \approx 1.09$ \citep{2012ApJ...755..164A}, and $k \approx 1.27$ \citep{2020ApJ...894...88A}.

Both the FRC and F$\gamma$C have received considerable theoretical attention. Most models attribute the mild superlinearity of the F$\gamma$C to galaxies being imperfect and variable proton calorimeters, such that only a fraction of the CR protons interact with the ISM and produce $\gamma$-rays\footnote{In the dense gas of the galactic disc, streaming losses are small compared to collisional losses, hence we neglect them \citep{2019MNRAS.488.3716C}.}, but  this fraction systematically increases with galaxy surface density. This, in turn, correlates with star formation rate and FIR luminosity, giving rise to a slightly increasing $\gamma$-ray to FIR ratio at high luminosity \citep[e.g.,][]{Lacki11a, Werhahn21b, 2021MNRAS.502.1312C, Crocker21b}.

Models to explain the FRC can be broken into two broad categories. One is based on the idea that galaxies are good CR electron calorimeters, and that synchrotron emission is the dominant electron loss mechanism \citep[e.g.,][]{1989A&A...218...67V}. While this would naturally explain a tight, linear FRC, in order to avoid producing too much synchrotron radiation compared to what is observed, this model requires that supernovae accelerate CR electrons much less efficiently than observations of individual Milky Way supernova remnants suggest. It also does not explain why contributions from secondary electrons, or, conversely, bremsstrahlung and ionisation losses, do not cause a deviation from linearity in dense starburst galaxies where these processes should become important. Perhaps most significantly, this model predicts the wrong radio spectral index \citep{2006ApJ...645..186T}: a CR electron population dominated by synchrotron (or inverse Compton) losses should produce synchrotron radiation with a spectral index $\alpha\approx 1.1$, but observations yield $\alpha \approx 0.6-0.8$ \citep{1982A&A...116..164G, 2000A&A...354..423L, 2010MNRAS.401L..53I}.

In a variant of this model, \citet{2013A&A...556A.142S} argue that galaxies do not radiate most of their CR electron luminosity as synchrotron emission, but that the fraction that goes into synchrotron versus inverse Compton radiation is maintained at nearly constant levels by the action of the turbulent dynamo \citep{1996A&A...306..677L, 2006ApJ...645..186T, 2010ApJ...717....1L}, which maintains a nearly constant Alfv\'en Mach number in the ISM. While this resolves the problem of requiring very low acceleration efficiency, it does not resolve the additional challenge of explaining the absence of curvature from secondaries and ionisation / bremsstrahlung losses, or of reproducing the radio spectral index.

The other, alternative approach is to explain the linearity of the FRC as arising from several accidental cancellations \citep[e.g.,][]{2010ApJ...717....1L, 2010ApJ...717..196L, Werhahn21c}, a phenomenon often referred to as a ``conspiracy'' in the literature. In high-density systems such as starbursts, increased synchrotron emission from secondary CR electrons generated by proton-ISM interactions can plausibly be cancelled by a reduction of the fraction of all CR electron power going into synchrotron emission as a result of increasing bremsstrahlung, Coulomb, and ionisation losses. This cancellation may seem coincidental \citep{2010ApJ...717....1L, 2013ASSP...34..283T}, but can in fact be explained by the underlying physics which requires that the loss rates scale linearly with the ISM density for all these mechanisms.
Conversely, in low-density galaxies, a significant fraction of CR electrons escape, depressing the radio luminosity \citep{2003ApJ...586..794B}, but this is compensated by escape of optical and UV photons from starlight, which depresses the FIR luminosity by similar amounts.

While these theoretical efforts are encouraging, they remain incomplete. In particular, at present we lack a model that can simultaneously and self-consistently explain (1) the slope and normalisation of the F$\gamma$C, (2) the slope, normalisation, and linearity of the FRC, and (3) the radio spectral index, all across the full range of star formation rate probed by observations. \citet{2010ApJ...717....1L, 2010ApJ...717..196L} have come closest to achieving this, however these works still predict somewhat harder spectral indices than observed.
Our goal in this paper is to supply a model that satisfies all three requirements, and use it to pick apart the physical origins of these correlations, and make predictions for additional covariances between non-thermal emission and other galaxy properties that can be used as further tests of the model.

To this end, we make use of the \textsc{CONGRuENTS} (COsmic-ray, Neutrino, Gamma-ray and Radio Non-Thermal Spectra) model developed in \citet[hereafter \citetalias{roth+2023}]{roth+2023}. This model features a number of improvements on prior work. First, it uses a combination of observed correlations and physical principles such as hydrostatic balance and dynamo action to derive the properties of galaxy ISMs, yielding predictions for the gas, magnetic fields, and radiation fields with which CRs interact that are both more realistic and have fewer free parameters than earlier models. Second, rather than simply adopting a fixed CR diffusion coefficient, it uses a physical model for CR transport and escape \citep{2020MNRAS.493.2817K}, and thus offers a self-consistent prediction for CR calorimetry and secondary production as a function of galaxy properties. Third, \textsc{CONGRuENTS} provides a full kinetic treatment of electron energy transport, properly capturing non-local energy jumps caused by bremsstrahlung and inverse Compton scattering in the Klein-Nishina regime. In \citetalias{roth+2023} we show that this treatment yields accurate predictions for the full non-thermal spectra of most nearby galaxies for which both radio and $\gamma$-ray data are available; \citet{2021Natur.597..341R} show that an earlier version of the same model also successfully explains the diffuse extragalactic $\gamma$-ray background as due to the cosmologically integrated emission of star-forming galaxies (see also \citet{2007ApJ...654..219T}).

In this paper we apply \textsc{CONGRuENTS} to a large sample of observed galaxies with measured stellar masses, star formation rates, and radii -- ensuring that our input galaxy parameters are realistically correlated with one another -- and predict non-thermal radio and $\gamma$-ray spectra for the full sample. We describe our method, and the galaxy sample to which we apply it, in \autoref{sec:methods}. We then compare this synthetic sample to observations in \autoref{sec:results}, demonstrating that we successfully reproduce all the observed correlations listed above. In \autoref{sec:discussion} we analyse these results in order to elucidate the physical origins of the various correlations, and we introduce new testable predictions from our model. Finally, in \autoref{sec:conclusion} we summarise our results and discuss prospects for future work.

\section{Methods}
\label{sec:methods}

In order to explain correlations between thermal and non-thermal emission and understand their origin, we first construct a large set of predicted non-thermal spectra. We describe the method by which we compute these spectra for a galaxy with a specified stellar mass, star formation rate, and radius in \autoref{ssec:review}, and we describe the sample observed galaxy properties that we use as inputs when constructing our synthetic catalogue in \autoref{ssec:candels}. We discuss how this catalogue compares to some of the observational samples of non-thermal emission to which we will compare in \autoref{ssec:sampling}.

\subsection{Review of model and \texorpdfstring{\citetalias{roth+2023}}{CONGRuENTS I}}
\label{ssec:review}

In this second paper of the series, we derive  results using \textsc{CONGRuENTS}, presented in \citetalias{roth+2023}, and will only provide a brief overview of the solution method in this paper. We refer the reader to \citetalias{roth+2023}
for a full description of the method, along with a number of validation tests showing that it makes accurate predictions for the spectra of well-observed local galaxies.

The fundamental inputs to a \textsc{CONGRuENTS} calculation are the stellar mass, star formation rate, and optical radius of a galaxy; we derive all our predictions from these quantities. In order to do so, we model galaxies in their restframes
as a two zone slab configuration. These two zones model i) a hydrostatic disc of mostly neutral gas and ii) an ionised halo, with both pervaded by a plane parallel interstellar radiation field (ISRF). 
The ISRF,
broadly following the works by \citet{2011piim.book.....D}, is made up of a (redshift dependent) black body cosmic microwave background; dilute black bodies with temperatures of 3000 and 4000 K produced by the old stellar population; a hot 7000 K dilute black body and a UV field \citep{2011piim.book.....D} driven by newly-formed stars; and a far-infrared radiation field produced by dust warmed by star formation. The intensities of all these components (except the CMB) are functions of the galactic stellar mass, star formation rate, and galactic disc area, with the scaling between these quantities and the radiation field components obtained from a fit to a sample of spiral and irregular galaxies in the DustPedia sample \citep{2019A&A...624A..80N}. These same fits predict the FIR luminosity of galaxies, where we assume that the entire dust reprocessed starlight is emitted in the band 8-1000$ \mu$m[\footnote{Though we caution the reader that a range of FIR bands are used in the literature; cf. \citealt{2015A&A...573A..45M}}]. FIR emission deserves special attention in this work since it is a key observable; our fitted relation is
\begin{equation}
    \frac{L_\mathrm{FIR}}{\mathrm{L}_\odot} = 5.13\times 10^9 \left(\frac{\dot{M}_{\ast}}{\mathrm{M}_\odot\,\mathrm{yr}^{-1}}\right)^{1.10},
    \label{eq:LFIR}
\end{equation}
and we use this throughout the paper whenever we require predictions of the FIR luminosities of our synthetic galaxies. We provide explicit formulae for the remaining components of the radiation field in \citetalias{roth+2023}. 

We derive the magnetic field in the disc and halo of each galaxy as a function of the gas velocity dispersion, the density of the interstellar medium and the Alfv\'en Mach number, as outlined in \citet{roth+2023}. While the former two values are determined for each galaxy, we assume a universal Alfv\'en Mach number $\mathcal{M}_{\rm A} = 2$ for all galaxies, based on the expectations of dynamo theory. We refer the interested reader to Section 2 in the Supplementary Information of \citet{2021Natur.597..341R} for a detailed discussion on the appropriateness of this choice and the impact of alternative choices on model outputs.

To compute the CR population in a CONGRuENTS model, we assume that 10\% of
a core collapse supernova kinetic energy ($10^{50}$erg per SN) is injected in CR protons \citep{1995STIN...9622970W, 2013A&A...553A..34D}, supplemented with a further 2\% \citep{2010ApJ...717....1L} in primary CR electrons ($2 \times 10^{49}$erg per SN); we discuss the implications of these choices for our results in \autoref{sssec:secondaries}. CRs are then subject to a number of loss processes. For CR protons (CRp) we consider diffusive escape and $pp$-collisions with the ISM in the disc. Following the work of \citet{2020MNRAS.493.2817K}, this allows us to directly calculate the fraction of CRp energy that is converted to $\gamma$-rays and secondary particles, known as the calorimetry fraction \citep{2006ApJ...645..186T, 2021Natur.597..341R}. For CR electrons and positrons we have to consider a wider range of loss processes. In the disc we consider synchrotron, inverse Compton (on ISRF targets), and bremsstrahlung emission, ionisation losses, and diffusive escape into the halo. In the halo, we consider the same range of processes, except that, since the medium is fully ionised, ionisation losses are replaced by Coulomb losses, though here the low density ensures that both these mechanisms and bremsstrahlung are mostly negligible.

Since some of the loss mechanisms for CR electrons cause large jumps in CR electron energy with each interaction, we determine the steady-state CR electron and positron spectra in both the disc and the halo by numerical solution of the full kinetic equation, where we treat synchrotron and ionisation losses as gradual loss processes and inverse Compton and bremsstrahlung losses as potentially catastrophic (i.e., a large fraction of energy can be lost in a single interaction). We then proceed to compute the emission in synchrotron, inverse Compton and bremsstrahlung, and attenuate the emitted spectrum by the free-free opacity at radio frequencies, and by the internal $\gamma\gamma$ pair-production opacity at $\gamma$-ray energies. Finally we add the corresponding free-free emission to the spectrum. 

\begin{figure*}
	\includegraphics[width=\textwidth]{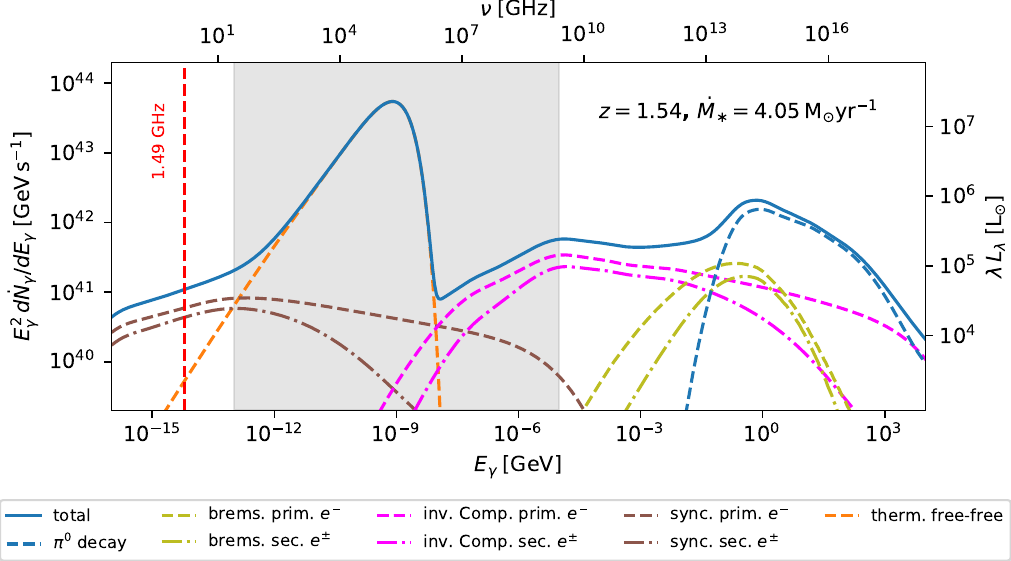}
    \caption{A \textsc{CONGRuENTS} prediction for the broadband non-thermal (and thermal free-free emission) spectrum of a galaxy drawn randomly from our sample; the galaxy shown has a star formation rate $\dot{M}_{\ast} = 4.05$ M$_\odot$ yr$^{-1}$, a stellar mass $M_{\ast} = 2.4\times 10^{9}$ M$_\odot$, a half-light radius $R_{\rm e} = 3.14$ kpc and is at redshift $z=1.54$, but the frequencies and photon energies we plot on the horizontal axis are the emitted-frame values, not observed-frame values. The solid blue line shows the total spectrum, and other line colours and styles indicate the contributions from various mechanisms and particle populations, as indicated in the legend; although not shown, the model output can also be decomposed into disc and halo emission components. The grey band gives an approximate range of energies where thermal emission is expected to be dominant over non-thermal emission, so the non-thermal component is obscured. The target spectrum for inverse Compton emission comprises the components of the ISRF, as described in the text, however it does not include thermal free-free emission whose overall contribution to the ISRF is negligible.}
    \label{fig:singlespec}
\end{figure*}

We show in \autoref{fig:singlespec} an example of the (restframe) emission spectrum $E_\gamma^2 (d\dot{N}_\gamma/dE_\gamma)$ predicted by \textsc{CONGRuENTS} for a random galaxy in the sample we describe in \autoref{ssec:candels}; here $E_\gamma$ is the photon energy and $d\dot{N}_\gamma/dE_\gamma$ is the number of photons emitted per unit time per unit energy. As the figure shows, the \textsc{CONGRuENTS} model produces a full spectrum at radio and $\gamma$-ray energies, decomposed into the parts of the emission driven by each mechanism and each population of particles; although not shown in the figure to avoid clutter, the emission can also be decomposed into disc and halo components.
These results directly give us radio fluxes at commonly-observed frequencies such as $1.49$ GHz. We obtain total $\gamma$-ray luminosities, which are required when comparing to observations from \textit{Fermi}, by integrating the spectra between 0.1 and 100 GeV in the source frame. Similarly, in order to compare to radio observations we require radio spectral indices. We compute these between any given pair of frequencies $\nu_{1}$ and $\nu_{2}$ (again measured in the emitted frame) using the logarithmic slope of the flux density,
\begin{equation}
\label{eq:spectralindex}
    \alpha_{\nu_1}^{\nu_2} = -\frac{\log(S_{\nu_2}/S_{\nu_1})}{\log(\nu_2/\nu_1)},
\end{equation}
where $S_{\nu_2}$ is the flux density at frequency $\nu_2$, and similarly for $S_{\nu_1}$. We explicitly use the convention $S_{\nu} \propto \nu^{-\alpha}$. Where we indicate only a single frequency (e.g., $\alpha_{1.49\,\mathrm{GHz}}$), we compute this using frequencies $\nu_1$ and $\nu_2$ on either side of the target frequency, spaced so that $\log\left( \nu_{2}/\nu_{1}\right) = 0.05$.

\subsection{The CANDELS sample}
\label{ssec:candels}

Because the non-thermal emission produced by a galaxy depends on its stellar mass (gravity and starlight), star formation rate (gas content, supernova rate, starlight), and radius, when producing a synthetic catalogue it is important to ensure that these parameters not only fall in realistic ranges, but that they are realistically correlated with one another. Thus, for example, there is a well-known correlation between stellar mass and star formation rate known as the star-forming main sequence (SFMS), and the fact that star-forming galaxies fall along this sequence may be important for shaping the distribution of their non-thermal emission properties. For this reason we wish to generate our synthetic non-thermal emission catalogue starting from a set of real, observed galaxy stellar properties.

For this purpose we use the galaxies observed as part of the Cosmic Assembly Near-infrared Deep Extragalactic Legacy Survey (CANDELS; \citealt{2011ApJS..197...35G}). This survey was designed to sample galaxy evolution at redshifts $z \sim 1.5$ to $\sim$8, with complete sampling of the population down to stellar masses of $10^{9}$ M$_{\odot}$ up to $z \approx 2$, and a lower limit beyond $z \approx 2$ down to the knee of the ultraviolet luminosity function. We take our sample from the part of CANDELS that overlaps the GOODS-S field, which covers an area of $173 \ {\rm arcmin^{2}}$; \citet{2012ApJS..203...24V} analyse the structural properties of galaxies in this field, and derive their stellar masses $M_{\ast}$, star formation rates $\dot{M}_{\ast}$, and effective radii $R_{\rm e}$ corrected to 5000 \AA~\citep{2014ApJ...788...28V}, supplemented by a redshift $z$. Using the flags provided by \citet{2012ApJS..203...24V}, we filter these data to remove galaxies that are likely to house an active galactic nucleus (AGN), for which the fit to the brightness profile is poor, or for which the redshift or star-formation rate is unreliable. We also remove objects that are otherwise flagged as suspicious in the catalogue. Out of an initial 34930 galaxies, this leaves a total of 20346 galaxies that are suitable for processing using our model.

This provides us with a uniform, realistic set of input stellar parameters for \textsc{CONGRuENTS} modeling. We use these stellar parameters as inputs to the procedure described in \autoref{ssec:review}, producing predicted non-thermal emission for the full sample.

\subsection{Observational comparison samples and the importance of sampling}
\label{ssec:sampling}

CANDELS offers us a well-curated, deep sample of galaxies that provides excellent coverage across the SFMS, and thus captures the dominant types of star-forming galaxies and the dominant sources of star formation-powered non-thermal emission. However, this coverage does not necessarily match that of the surveys of non-thermal emission to which we will be comparing below, and these differences are important to keep in mind. 

Most significantly, while CANDELS is deep, it is not wide, and its limited field of view means that it has limited coverage of rare galaxy types at redshifts well below $\sim 0.5$, where it is susceptible to cosmic variance. Thus for example CANDELS contains few low-redshift starburst galaxies like M82 or Arp 220, because these are rare and thus are unlikely to occur within the small CANDELS field of view. We expect to miss a significant fraction of passive galaxies at high redshift for the same reason, though these are presumably not important as non-thermal emitters driven by star formation.

By contrast, many observational surveys of non-thermal radio emission are restricted to bright, low-redshift sources that are readily observable \citep[e.g.,][]{2003ApJ...586..794B, 2017ApJ...847..136B}, with limited effort to construct uniform samples across the SFMS. These surveys tend to over-sample bright galaxies such as Arp 220 relative to their true number fraction because they are easier to observe, a bias in exactly the opposite direction as CANDELS. Due to this bias, in this paper we will refrain from comparing the redshift evolution of our model results with observational results, as carefully addressing this in detail would require that we understand the selection used in the observations and its evolution with redshift, then weight the CANDELS sample to match it; doing so is beyond the scope of this paper.

The situation is even more 
problematic
for $\gamma$-ray observations, where we are restricted to a handful of the closest and brightest galaxies that have been observed with the \textit{Fermi} satellite \citep[e.g.,][]{2012ApJ...755..164A}. Where we require $\gamma$-ray data, we use the radio, FIR and $\gamma$-ray luminosities tabulated in \citet{2020A&A...641A.147K, 2022A&A...657A..49K}, removing galaxies from the sample that are known to host AGN. This leaves a $\gamma$-ray observed sample of a mere 10 galaxies, including the Milky Way. This collection includes two dwarf satellite galaxies (the SMC and LMC) and two galaxies far off the SFMS (M31 and Arp 220); neither of these categories is well-represented in CANDELS. In addition, a number of these systems are so close to our own Galaxy that they are observed as extended sources in the sky, which introduces significant uncertainties in both their radio and $\gamma$-ray luminosities due to the need to model and subtract the Galactic foreground 
\citep[e.g.,][]{2018MNRAS.480.2743F}. The Milky Way's own total non-thermal radio and $\gamma$-ray luminosities are also substantially uncertain, with different analyses producing contradictory results \citep[e.g.,][]{2010ApJ...722L..58S, 2012ApJ...750....3A}.

In \citetalias{roth+2023} we present a bespoke analysis of these local galaxies' non-thermal spectra and show that we can reproduce most of them reasonably well, but here, where our goal is to understand population-level correlations, such an approach is not available. For this reasons we will continue to use our predicted non-thermal emission for CANDELS galaxies as the basis for our analysis, but we caution that we expect some differences between the trends in this population and those found in the ten-galaxy local sample simply because of differences in the underlying galaxy populations.

\section{Results}
\label{sec:results}

\subsection{Radio emission}

We now  present the results of our calculation. We will first discuss radio emission and present the FRC we derive for our galaxy sample and put this into context with observational results (\autoref{sssec:FRC}). We then  show the radio spectral indices we derive for our sample, in terms of the total and the decomposed non-thermal (synchrotron only) spectral index after removing the contribution from free-free emission (\autoref{sssec:radio_index}).

\subsubsection{The FIR-radio correlation}
\label{sssec:FRC}

\begin{figure*}
	\includegraphics[width=\textwidth]{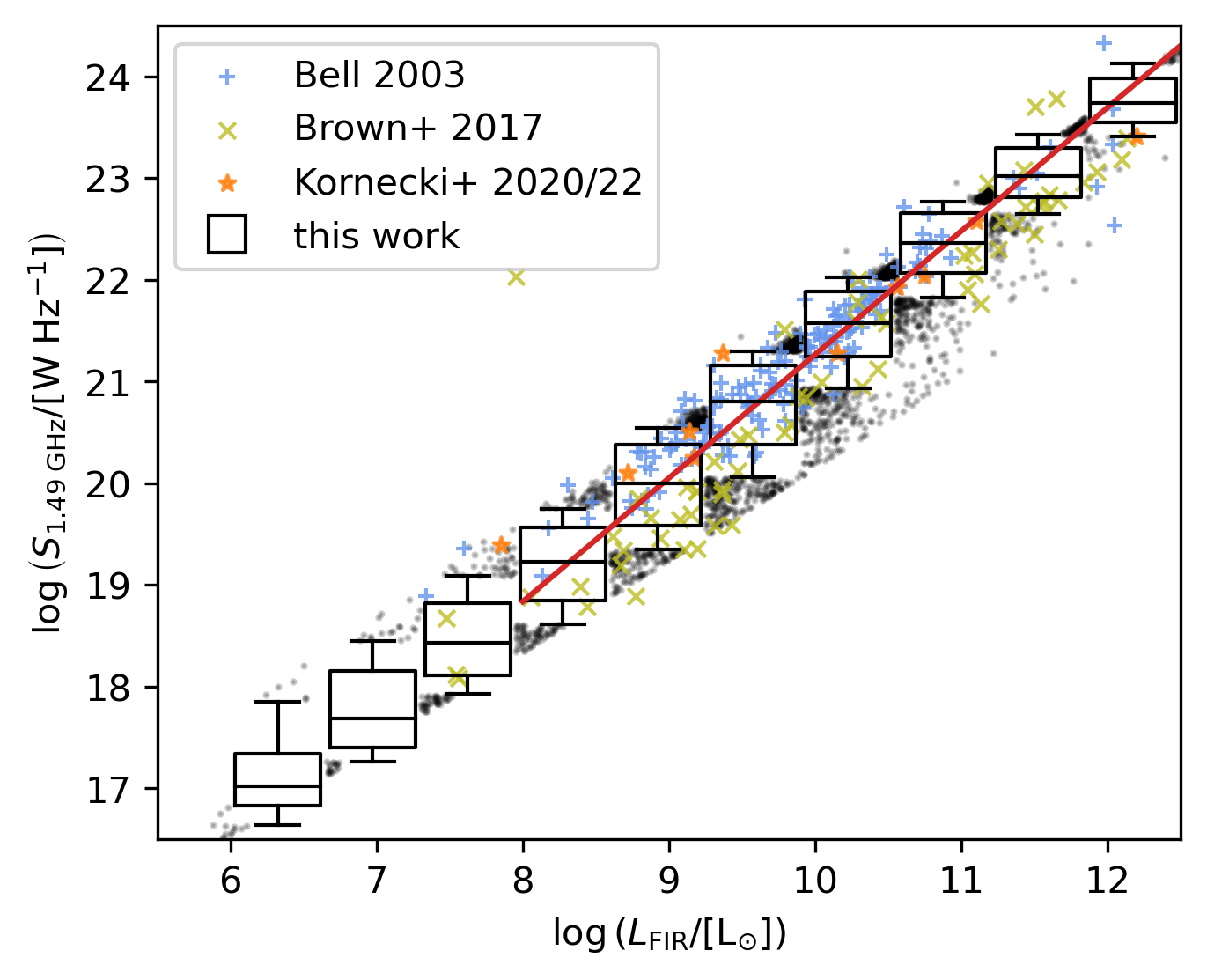}
    \caption{Boxes and whiskers show our computed FIR luminosity and 1.49 GHz flux density for the CANDELS sample; boxes show the mean (central line) and $16^{\rm th}$ to $84^{\rm th}$ percentile range of $S_\mathrm{1.49\,GHz}$, and whiskers show the $5^{\rm th}$ to $95^{\rm th}$ percentile range, computed over the indicated intervals of $L_\mathrm{FIR}$. Black points correspond to our predictions for individual CANDELS galaxies that scatter beyond the range indicated by the whisker in each SFR bin. The red line is a power-law fit to the data points with $10^{8} L_\odot < L_{\rm FIR} < 10^{13} L_\odot$. Points marked with + are the observed galaxies from the sample of \citet{2003ApJ...586..794B}, points marked with $\times$ are observations from \citet{2017ApJ...847..136B}, and points marked with $\star$ represent galaxies taken from \citet{2020A&A...641A.147K,2022A&A...657A..49K}.}
    \label{fig:FIRradio}
\end{figure*}

To derive the FRC we predict for the CANDELS sample, we combine the 1.49 GHz specific luminosity predicted by \textsc{CONGRuENTS} with the FIR luminosity derived from \autoref{eq:LFIR}.
We plot the result in \autoref{fig:FIRradio}; in this figure, boxes show the mean and $16^{\rm th}$ and $84^{\rm th}$ percentiles of the 1.49 GHz radio luminosity, and whiskers the $5^{\rm th}$ and $95^{\rm th}$ percentiles, computed over the indicated intervals in FIR luminosity. A least-squares fit of the distribution for galaxies with $8 < \log(L_\mathrm{FIR}/\mathrm{L}_\odot) < 13$ (roughly the FIR luminosity range spanned by the observational samples to which we will compare) yields a super-linear FIR radio correlation
\begin{equation}
    \log{ \left( \frac{S_{\rm 1.49 GHz}}{\rm W\, Hz^{-1}} \right) } = 1.2 \log\left( \frac{L_{\rm FIR}}{\rm L_\odot}\right) + 9.1,
\end{equation}
which we show as the red line in \autoref{fig:FIRradio}. The results we obtain are qualitatively consistent with the distribution of observed galaxies \citep{2003ApJ...586..794B, 2017ApJ...847..136B, 2020A&A...641A.147K, 2022A&A...657A..49K}, where we have taken the FIR luminosities as tabulated in \citet{2020A&A...641A.147K, 2022A&A...657A..49K, 2003ApJ...586..794B} and use our conversion from \autoref{eq:LFIR} to convert the star-formation rates in \citet{2017ApJ...847..136B}, though it is clear that there are systematic uncertainties within the data themselves. For example, it is evident that the star-formation rates, and by extension the FIR luminosities, that \citet{2017ApJ...847..136B} obtain for dwarf galaxies are categorically higher than the results in \citet{2003ApJ...586..794B} and other prior literature, a discrepancy that \citeauthor{2017ApJ...847..136B} note themselves. Our model galaxies broadly cover the range of both samples but are somewhat closer to the results obtained by \citet{2003ApJ...586..794B}. The apparent upper and lower bounds for the radio emission as computed by the model are set by the 100\% calorimetry and by the contribution from thermal free-free emission, respectively.

It is also worth noting that, viewed in isolation, the \citet{2020A&A...641A.147K, 2022A&A...657A..49K} sample (orange stars in \autoref{fig:FIRradio}) shows a significantly shallower FRC slope than the other two data sets or than our model predictions. This is likely because of differences in the galaxy population it samples, as discussed in \autoref{ssec:sampling}. This point will be significant when we examine the radio-$\gamma$-ray correlation below.

\begin{figure}
	\includegraphics[width=\columnwidth]{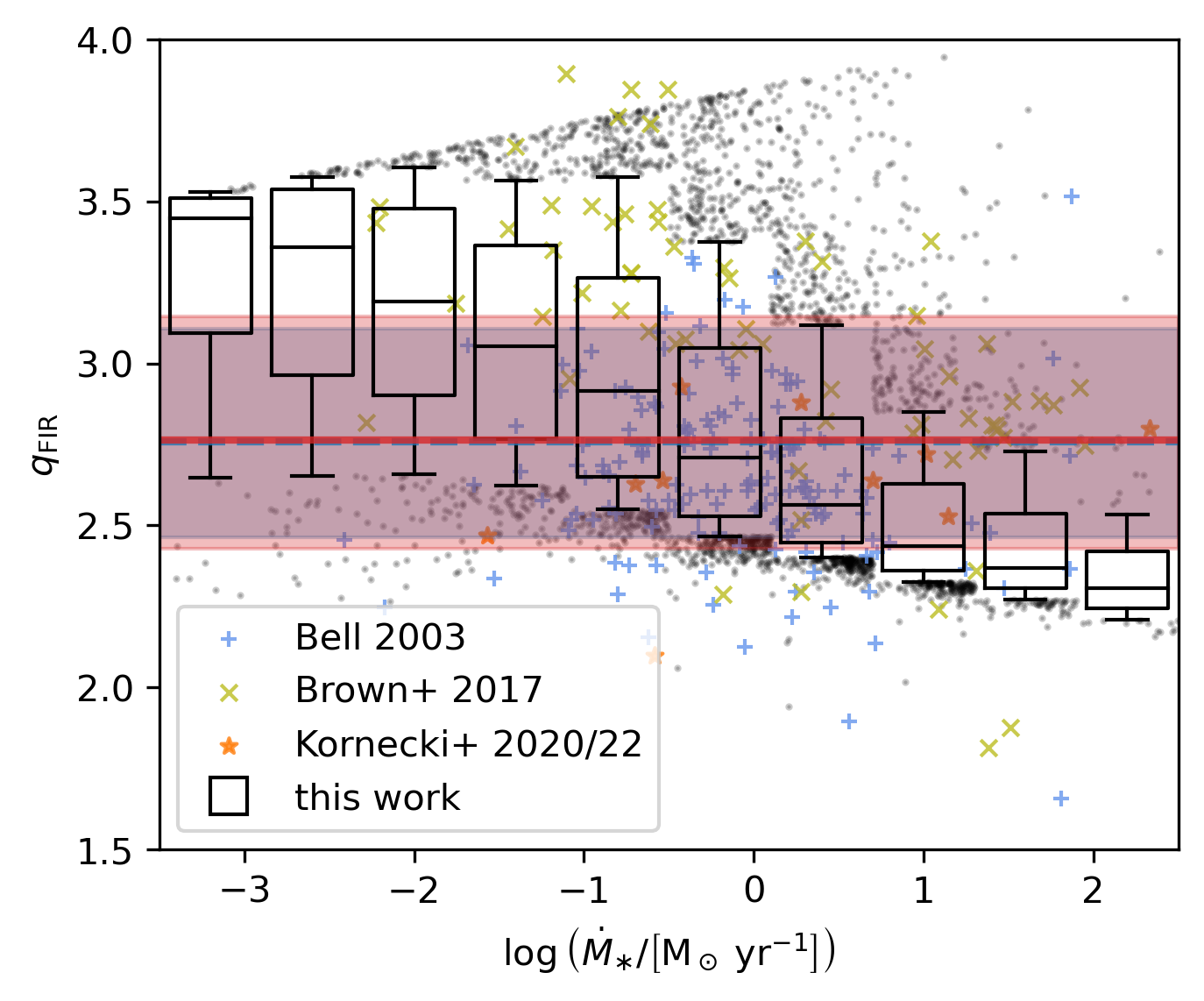}
    \caption{Log ratio of the infrared and 1.49 GHz luminosities $q_{\rm FIR}$ versus star formation rate; $q_\mathrm{FIR}$ is as defined by \citet{1985ApJ...298L...7H}, and we take $L_{\rm FIR}$ to be the integrated luminosity between 8 and 1000 $\mu$m. Boxes, whiskers, black points and symbols have the same meaning as in \autoref{fig:FIRradio}.
    The red horizontal line shows the mean of all model-predicted galaxies, while the red shaded band shows the 16th to 84th percentile range around this mean. The (almost obscured) blue dashed line represents the mean for the three observational data sets combined \citep{2003ApJ...586..794B, 2017ApJ...847..136B, 2020A&A...641A.147K,2022A&A...657A..49K}, while the blue shaded band shows their 16th to 84th percentile range.}
    \label{fig:qirSFR}
\end{figure}

In \autoref{fig:qirSFR} we show the FIR-radio correlation in the log-ratio parameterisation \citep{1985ApJ...298L...7H, 2021A&A...647A.123D} given by
\begin{equation}
q_{\rm FIR} = \log\left( \frac{L_{\rm FIR}}{3.75 \times 10^{12} \; {\rm W}} \right) - \log\left(\frac{S_{\rm 1.49 GHz}}{\rm W \; {Hz}^{-1}}\right).
\end{equation}
Boxes and whiskers have the same meaning as in \autoref{fig:FIRradio}. \autoref{fig:qirSFR} shows that, by this parameterisation as well, our predictions are qualitatively consistent with the locus in the SFR-$q_\mathrm{IR}$ plane found in observations. Quantitatively, for all the galaxies in our sample we obtain a mean $q_{\rm FIR}$ of $2.76_{-0.33}^{+0.38}$ (red band in \autoref{fig:qirSFR}, where the lower and upper limits are the 16th and 84th percentile respectively). This is in excellent agreement with the value $q_{\rm FIR} = 2.76_{-0.29}^{+0.35}$ (blue band in \autoref{fig:qirSFR}) we obtain by combining the galaxy samples from \citet{2003ApJ...586..794B}, \citet{2017ApJ...847..136B}, and data from references in \citet{2020A&A...641A.147K, 2022A&A...657A..49K}. Our result is also consistent with the mean values $q_{\rm FIR} = 2.62$ obtained by \citet{2001ApJ...554..803Y} and $q_{\rm FIR} = 2.63$ obtained by \citet{2015A&A...573A..45M} for $z \sim 0$ galaxies\footnote{For the purposes of this comparison we have converted these authors' reported $q_\mathrm{FIR}^{\prime}$ values, which we have denoted here with a prime and which refers to the integrated luminosity between 42-122$ \mu$m, to $q_\mathrm{FIR}$ as used in this work, which covers the band 8-1000$ \mu$m and which the authors refer to as $q_{\rm IR}$ instead, such that $q_{\rm FIR} = q_{\rm FIR}^{\prime} + 0.28$, following the conversion suggested in \citet{2015A&A...573A..45M}.}. \citet{2010A&A...518L..31I} obtain a somewhat lower but still consistent $q_{\rm FIR} = 2.40 \pm 0.24$. 

It is important to reiterate at this point that galaxy selection matters, and that the samples of galaxies used for these comparisons are very different from both one another and from CANDELS in terms of their completeness, redshift range, and sensitivity limits in infrared and radio luminosity. For example, the values $q_{\rm FIR} \approx 2.4$ obtained by \citet{2010A&A...518L..31I} is for a higher redshift sample than those of \citet{2003ApJ...586..794B}, \citet{2017ApJ...847..136B}, or \citet{2020A&A...641A.147K, 2022A&A...657A..49K}, all of whom target $z\sim 0$ galaxies. Qualitatively consistent with this trend, if we sub-divide the CANDELS sample, we find that the higher redshift, higher star-formation rate systems tends to have somewhat lower $q_\mathrm{IR}$ than the lower redshift, lower SFR part of the sample. However, we refrain from carrying out a detailed analysis of the redshift evolution of $q_{\rm FIR}$ \citep[e.g.][]{2010ApJ...714L.190S, 2010ApJS..186..341S, 2017A&A...602A...4D} because of its strong dependence on star-formation rate, and hence the assembled galaxy sample. Detailed comparison would therefore require careful modeling of the selection function and sensitivity limits of the survey to which we are comparing, which is beyond the scope of this work.

\begin{figure}
	\includegraphics[width=\columnwidth]{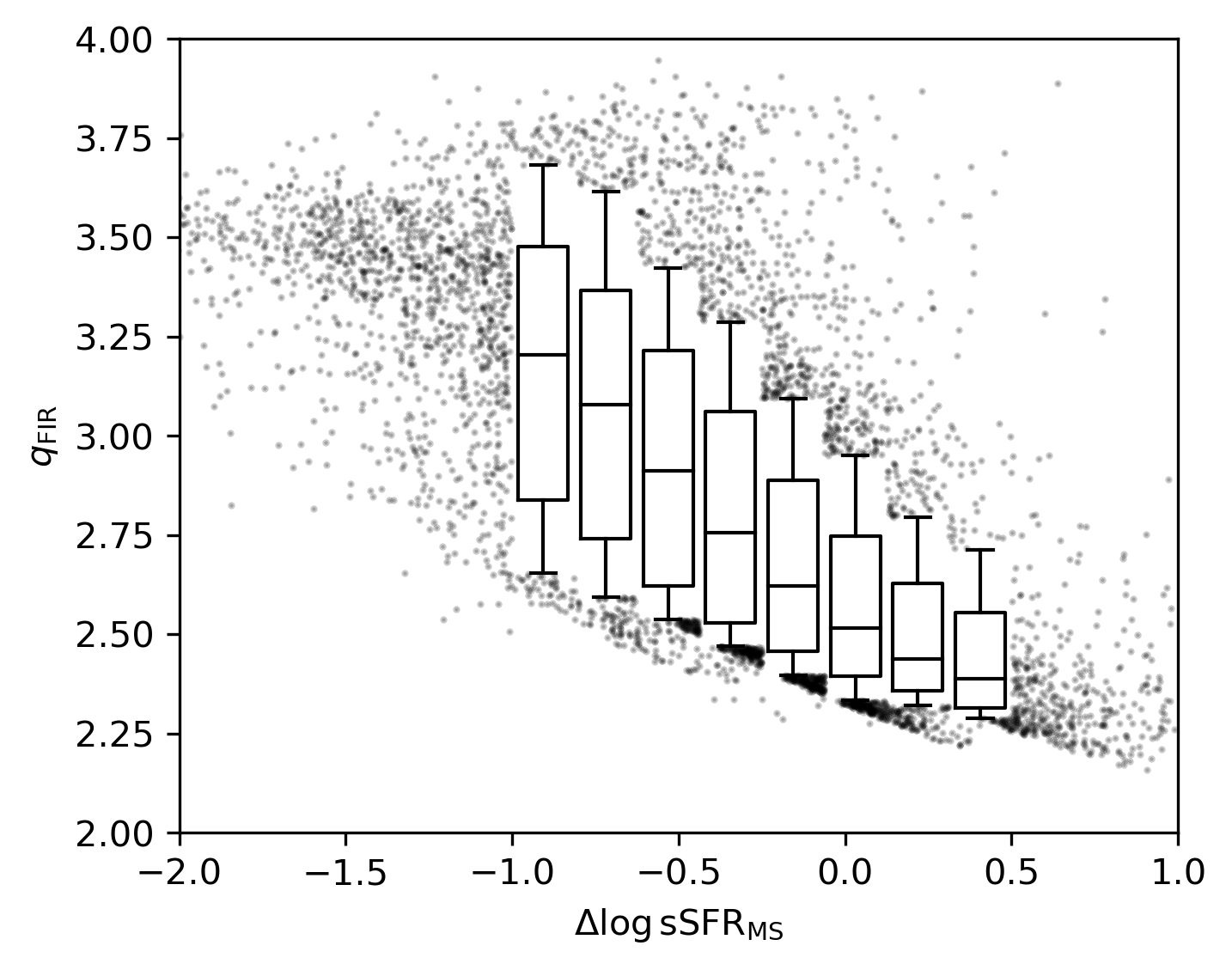}
    \caption{$q_{\rm FIR}$ as a function of distance from the star-forming main sequence $\Delta\log\mathrm{sSFR}_{\rm MS}$ (\autoref{eq:ssfr_ms}). Boxes, whiskers, and black points have the same meaning as in \autoref{fig:FIRradio}.}
    \label{fig:qirDeltasSFR}
\end{figure}

However, to shine some light on the problem of selecting an appropriate sample, we show in \autoref{fig:qirDeltasSFR} $q_{\rm FIR}$ as a function of distance from the star-forming main sequence $\Delta\log\mathrm{sSFR}_{\rm MS}$, defined as \citep{2015A&A...573A..45M}
\begin{equation}
    \label{eq:ssfr_ms}
    \Delta\log\mathrm{sSFR}_{\rm MS} = \log{\left(  \frac{\mathrm{sSFR}}{\mathrm{sSFR}_{\rm SFMS}\left( z, M_{\ast} \right)}\right)},
\end{equation}
where sSFR is the specific star formation rate of a galaxy (i.e., its star formation rate divided by its stellar mass, $\mathrm{sSFR} = \dot{M}_{\ast}/M_{\ast}$), and $\mathrm{sSFR}_{\rm SFMS}\left( z, M_{\ast} \right)$ is the specific star formation rate expected for a galaxy of mass $M_\ast$ at redshift $z$; we take this latter quantity from the parameterisation provided by \citet{2014ApJS..214...15S} and \citet{2018ApJ...853..179T}.  \autoref{fig:qirDeltasSFR} confirms that high values of $q_{\rm FIR}$, which are not commonly seen in samples of brighter galaxies \citep[e.g.,][]{2010A&A...518L..31I, 2015A&A...573A..45M}, occur mainly in dim galaxies that are far below the star-forming main sequence. Galaxies of this type are poorly-sampled in most published observational studies of the FRC, but are present in CANDELS (with the exception of stacking analyses such as \citet{2020ApJ...899...58L}, which lack the resolution to derive the structural parameters the model requires). In these systems we also might expect non-negligible millisecond pulsar emission that contributes to the radio synchrotron energy budget \citep{2021PhRvD.103h3017S,2022NatAs...6.1317C}, which \textsc{CONGRuENTS} currently does not model. Such a contribution would naturally depress $q_{\rm FIR}$ at low sSFR.

\subsubsection{The radio spectral index}
\label{sssec:radio_index}

\begin{figure}
	\includegraphics[width=\columnwidth]{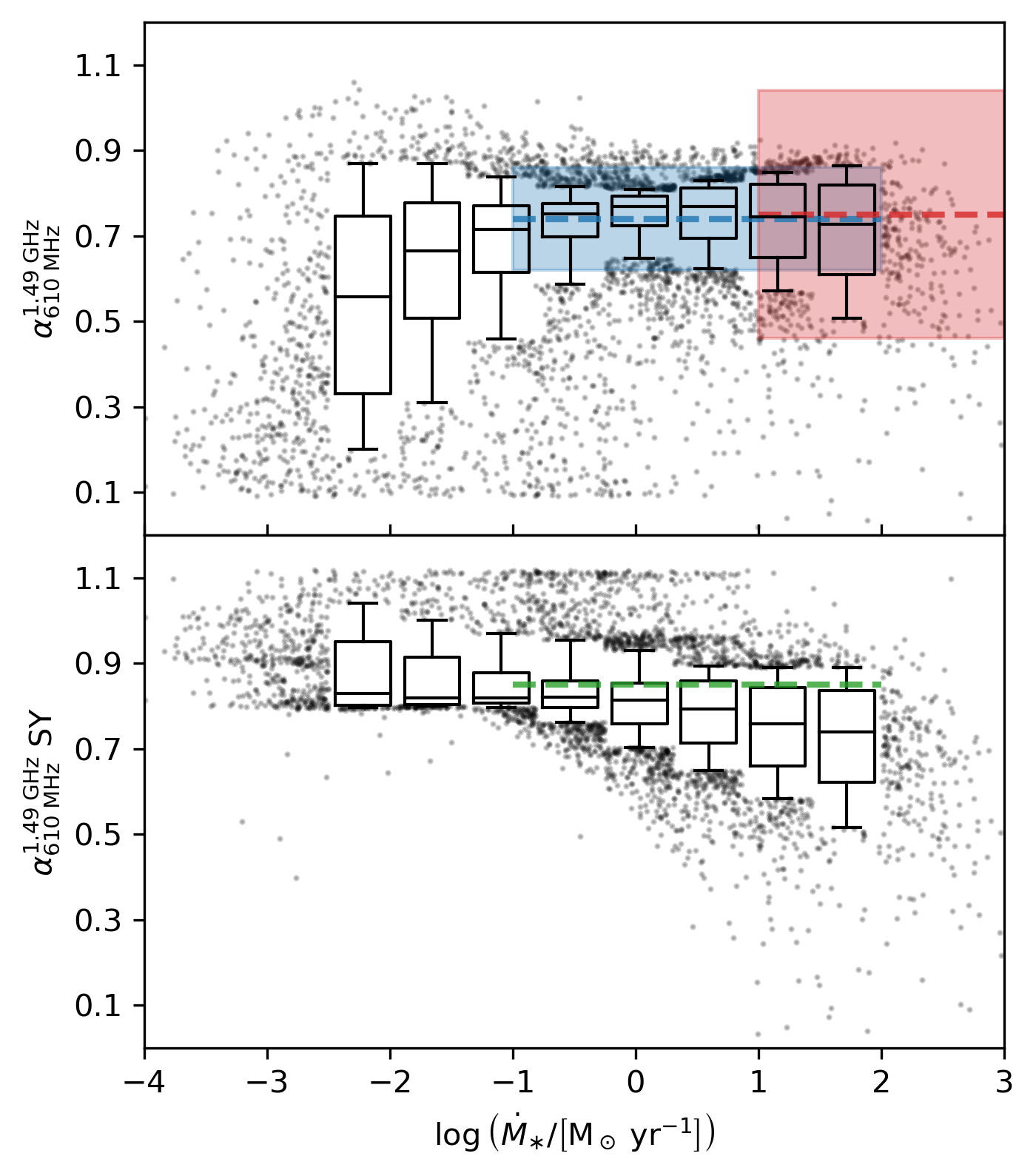}
    \caption{The upper panel shows the radio spectral between 610 MHz and 1.49 GHz as a function of star-formation rate. The lower panel shows the non-thermal spectral index after removing the contribution from free-free emission. Boxes, whiskers, and black points have the same meaning as in \autoref{fig:FIRradio}. The red dashed line in the upper panel shows the range of observed sub-mm galaxies in \citet{2010MNRAS.401L..53I}, with the horizontal extent of the line indicating the approximate rate of star formation rates covered by this study (see main text for details); the red band around this line indicates the 1$\sigma$ standard deviation in $\alpha_{\mathrm{610\,MHz}}^{\mathrm{1.49\,GHz}}$. The blue dashed line and band show the equivalent result obtained by \citet{1982A&A...116..164G} for bright spiral galaxies, while the green dashed line in the lower panel is the observational result derived by \citet{2000A&A...354..423L} for the spectral index of the synchrotron component alone, after removing the thermal free-free component, again shown over an indicative range of star formation rate.}
    \label{fig:alphaSFR}
\end{figure}

Following the common approach in observational work, we compute the radio spectral index $\alpha_{\rm 610 \, MHz}^{\rm 1.49 \, GHz}$ between 610 MHz and 1.49 GHz from the non-thermal emission we predict for the CANDELS sample using \autoref{eq:spectralindex}. We find a mean value and range $\alpha_{\rm 610 \, MHz}^{\rm 1.49 \, GHz} = 0.71_{-0.07}^{+0.08}$, where the upper and lower bound are the $16^{\rm th}$ and $84^{\rm th}$ percentiles. We show our predicted distribution of radio spectral index as a function of star formation rate in the upper panel in \autoref{fig:alphaSFR}. For comparison, in the lower panel we also show the spectral index $\alpha_{\rm 610 MHz, SY}^{\rm 1.49 GHz}$ that is produced by synchrotron emission alone, without the thermal free-free emission; for this case we obtain $\alpha_{\rm 610 MHz, SY}^{\rm 1.49 GHz} = 0.81_{-0.07}^{+0.06}$.

Our result for the total spectral index (including free-free effects) is in good agreement with the means obtained for sub-mm galaxies by \citet{2010MNRAS.401L..53I}, shown as the red dashed line and 1$\sigma$ band, and the sample of bright galaxies in \citet{1982A&A...116..164G} shown as the blue dashed line and 1$\sigma$ band. The green dashed line in the lower panel of \autoref{fig:alphaSFR} shows the result for the non-thermal spectral index after decomposition and removal of the free-free component as proposed by \citet{2000A&A...354..423L}, which is again consistent with the predictions of our model. For all three observational results, we plot the measurements over a range in star formation rate that is indicative of the sample used; however, these limits are not exact, since the authors of these studies do not fully describe their sample selection. We adopt $\dot{M}_{\ast} > 10 \, {\rm M_{\odot} \, yr^{-1}}$ as a rough lower limit for the star formation rate found in the sub-mm galaxy sample of \citet{2010MNRAS.401L..53I}, and $10^{-1} \, {\rm M_{\odot} \, yr^{-1}}$ to $10^{2} \, {\rm M_{\odot} \, yr^{-1}}$ to approximately characterise the \citet{1982A&A...116..164G} and \citet{2000A&A...354..423L} samples, both of which use spiral galaxies drawn from the NGC catalogue.

\subsection{\texorpdfstring{$\gamma$}{gamma}-ray emission}

We next examine two more observed correlations, the F$\gamma$C and the radio-$\gamma$ relation. We do this keeping in mind the caveats noted in \autoref{ssec:sampling} with respect to the sample from which the observational version of these correlations is constructed, and the differences in the selection between these samples and CANDELS.

\begin{figure}
	\includegraphics[width=\columnwidth]{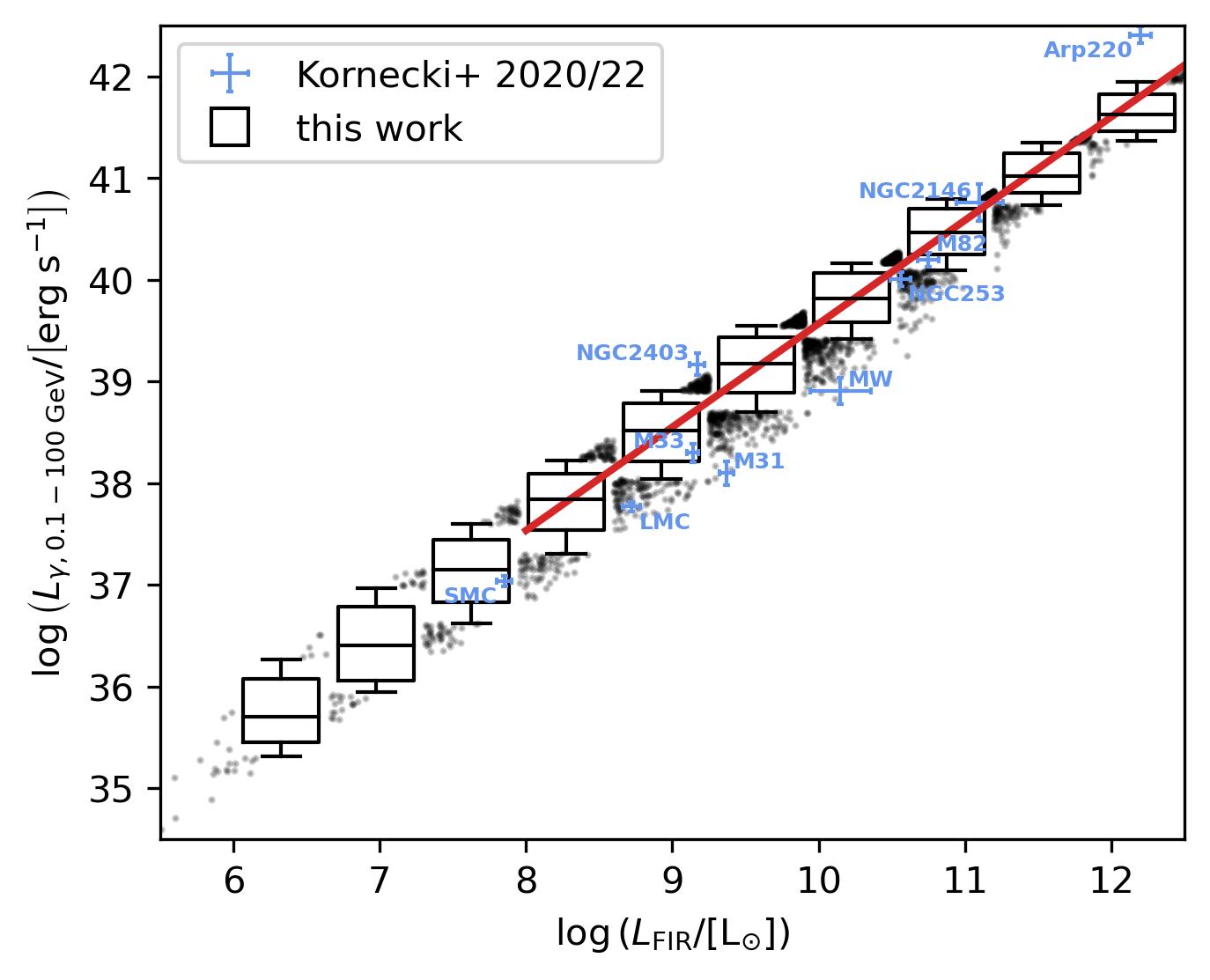}
    \caption{Integrated $\gamma$-ray luminosity over the energy range 0.1 - 100 GeV versus the FIR luminosity $L_{\rm FIR}$ between 8 and 1000 $\mu$m for our galaxy sample. Boxes, whiskers, and points have the same meaning as in \autoref{fig:FIRradio}. The red line is a power law fit to the data points in the range $10^{8}\,\mathrm{L_{\odot}} < L_{\rm FIR} < 10^{13}\,\mathrm{L_{\odot}}$. Blue data points with error bars show the observed galaxies from \citet{2020A&A...641A.147K,2022A&A...657A..49K}.}
    \label{fig:FIR_gamma}
\end{figure}
\begin{figure}
	\includegraphics[width=\columnwidth]{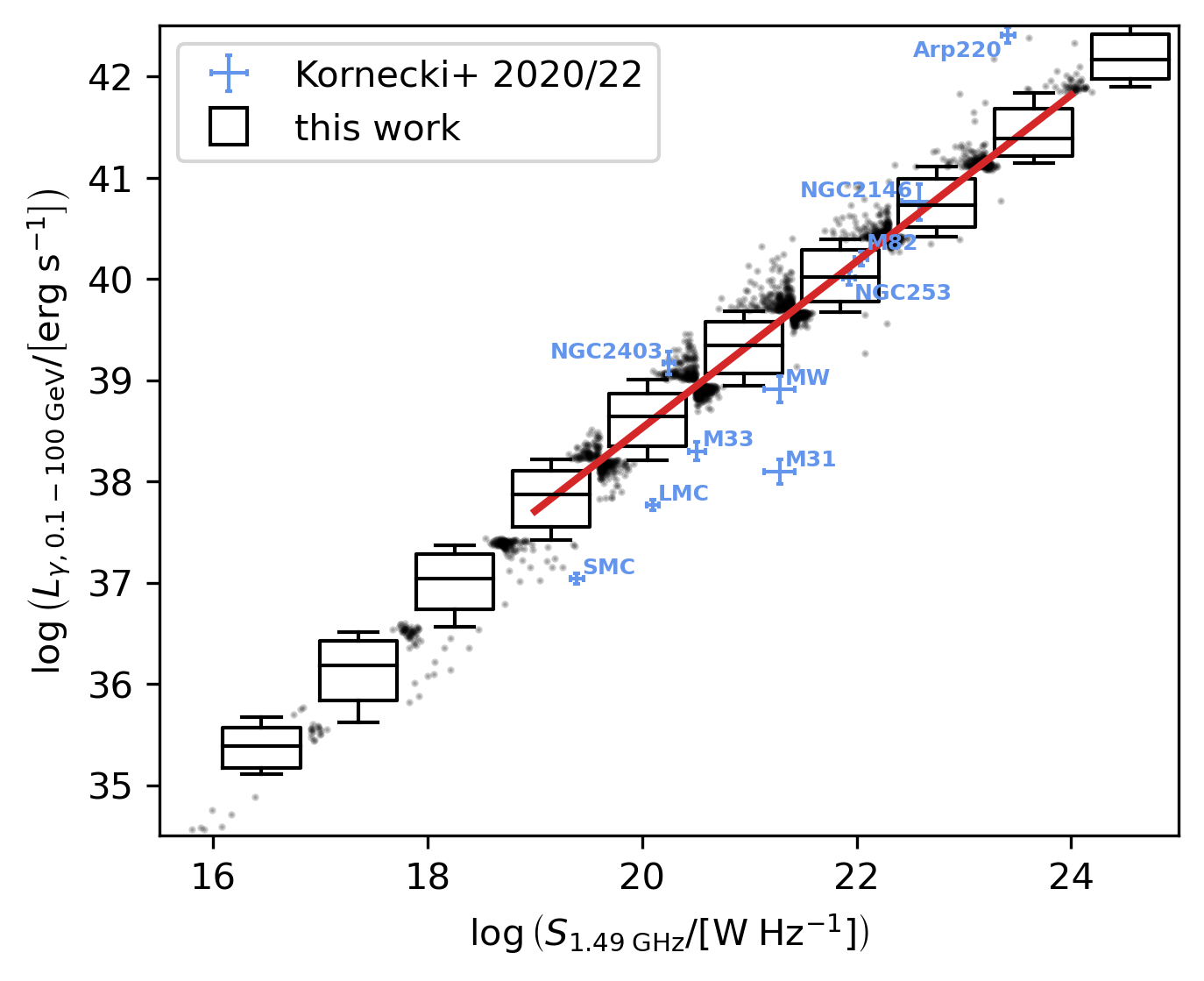}
    \caption{Integrated $\gamma$-ray luminosity between 0.1 - 100 GeV versus the 1.49 GHz flux for our galaxy sample. Boxes, whiskers, and black points have the same meaning as in \autoref{fig:FIRradio}. The red line is a power law fit to the data points in the range $10^{19}\,\mathrm{W \; Hz^{-1}} < S_{\rm 1.49 \; GHz} < 10^{24}\,\mathrm{W \; Hz}^{-1}$. Blue data points with error bars show the observed galaxies from \citet{2020A&A...641A.147K,2022A&A...657A..49K}.}
    \label{fig:gamma_radio}
\end{figure}

We show the result we obtain for the predicted F$\gamma$C of the CANDELS sample in \autoref{fig:FIR_gamma}. A least-squares fit of a power law functional form to our prediction for CANDELS yields
\begin{equation}
    \log\left(\frac{L_{\gamma,0.1-100\,\mathrm{GeV}}}{\mbox{erg s}^{-1}}\right) = 1.02 \log\left(\frac{L_\mathrm{FIR}}{\mathrm{L}_\odot}\right) + 29.4.
\end{equation}
We show this fit as the red line in \autoref{fig:FIR_gamma}. Both the predicted distribution for CANDELS and our fit to it are consistent with the best-fit slope of $1.09\pm 0.10$ that \citet{2012ApJ...755..164A} obtain, but our slope is significantly flatter than the $1.17 \pm 0.07$ obtained by \citet{2017PhRvD..96h3001L} and $1.21 \pm 0.07$ obtained by \citet{2020A&A...641A.147K}. We show in \autoref{fig:gamma_radio} the $\gamma$-ray-radio correlation we obtain for our model galaxies. A power law fit to the data yields
\begin{equation}
    \log\left(\frac{L_{ \gamma, {\rm 0.1-100 \; GeV}} }{ \rm erg \; s^{-1}}\right) = 0.82
    \log{\left(\frac{S_{\rm 1.49 \; GHz}} {\rm W \; Hz^{-1}}\right)} + 22.1.
\end{equation} 
Although there is general agreement between our predictions for CANDELS on the observed sample, our best-fit index of 0.82 is flatter than the relation derived by \citet{2012ApJ...755..164A}, who find an index of 1.10, and by \citet{2022A&A...657A..49K} who derive an index of 1.26.

The level of disagreement we find is unavoidable given the disagreements between the different observational data sets. As shown in \autoref{fig:FIRradio}, the \citet{2022A&A...657A..49K} sample has a significantly shallower FRC than either the
\citet{2003ApJ...586..794B} or \citet{2017ApJ...847..136B} data, and our model predictions lie much closer to the \citeauthor{2003ApJ...586..794B} and \citeauthor{2017ApJ...847..136B} results than to the \citeauthor{2022A&A...657A..49K} ones. The fact that our radio-$\gamma$-ray correlation is also shallower than the \citeauthor{2022A&A...657A..49K} one is just another manifestation of the same effect -- for example the fact that Arp 220 lies well to the left of our predicted radio-$\gamma$ correlation in \autoref{fig:gamma_radio} is mostly just because it is under-luminous in radio compared to other galaxies with similar SFRs (c.f.~\autoref{fig:FIRradio}, where Arp 220 is the data point from \citeauthor{2022A&A...657A..49K} at the highest FIR luminosity). Similarly, M31 is off our predicted radio-$\gamma$ correlation in the opposite direction mostly because it is very over-luminous in radio compared to other galaxies with similar FIR luminosities (again c.f.~\autoref{fig:FIRradio}, where M31 is the data point from \citeauthor{2022A&A...657A..49K} that is the farthest above the best-fit line). In general we observe that the galaxies in \citeauthor{2022A&A...657A..49K}'s local sample at moderate to high star-formation rates that lie close to the SFMS fall closer to our model predictions, while those that lie far above or below the SFMS are also noticeably further away from our modelled relation. 

Although our focus here is not on individual galaxies, it is worth noting that in \citetalias{roth+2023} we identify the specific properties of many of these galaxies that make them different from SFMS galaxies with similar star formation rates, and likely explain why the \citeauthor{2022A&A...657A..49K} sample shows systematic differences from both the larger \citeauthor{2003ApJ...586..794B} and \citeauthor{2017ApJ...847..136B} results and from our model predictions. Arp 220 for example has such a high visual extinction that its optical radius is a poor proxy for the radius of the nuclear starburst region where most of the star-formation is taking place. As a result of the small size of this region ($\sim 250$ pc) the free-free opacity of the ISM is much higher than it is for SFMS galaxies with similar star formation rates, which are typically much less compact and much less extincted; consequently Arp 220 is, in reality, much less radio-bright than might otherwise be expected. For M31, which is overluminous in radio compared to other galaxies at similar FIR luminosities, its non-thermal emission correlates much more closely with its old stellar population than with present-day star formation, and peaks in the quiescent bulge region, strongly suggesting that emission is dominated by old millisecond pulsars \citep{2021PhRvD.103h3017S, 2022MNRAS.516.4469Z} or, conceivably, thermonuclear supernovae, rather than by young stars and core collapse supernovae; by contrast neither pulsars nor thermonuclear supernovae from old stellar populations would be likely to dominate in a SFMS galaxy, which would have a much higher specific star formation rate. In a sample of only 10 galaxies, where these two galaxies anchor opposite ends of the data, these two effects alone are sufficient to noticeably flatten the best-fit FRC and steepen the radio-$\gamma$ slopes compared to what might be found for SFMS galaxies. This analysis highlights the importance of gathering a larger sample of $\gamma$-ray detected galaxies before drawing strong conclusions about population-level correlations in $\gamma$-ray emission.

A further possible contributor to the difference in slopes that such a sample would address is systematic differences between $z=0$ SFMS galaxies and the CANDELS sample, which is dominated by galaxies at $z \sim 1-3$. The latter have systematically higher gas fractions and star formation rates, and, probably most significantly from the standpoint of radio emission, different spectral shapes in their radiation fields (due to lack of older stellar population but a hotter CMB) and more turbulent interstellar media that beget stronger magnetic fields. These effects could well lead to subtle differences in the amount of electron power going into synchrotron versus other channels, and thus it is possible that there is a real systematic variation in the slopes of the IR-$\gamma$ and $\gamma$-radio correlations with redshift. Detecting such a difference will require a sample at $z=0$ that is systematically selected to sample the SFMS in the same way that CANDELS is at $z>0$.

\section{Discussion}
\label{sec:discussion}

We organise the discussion of our results as follows. 
First in \autoref{ssec:FRCexplanation} we explain what physical mechanisms underlie the FRC, and are responsible for determining its slope.
This is followed in \autoref{ssec:specindex} by a discussion of how the radio spectral index is shaped by the same mechanisms that produce the FRC, and explain how our model reproduces both the correlation and the spectral index distribution simultaneously.
Finally, in \autoref{ssec:fcal} we use the insight gained from the preceding discussion to predict a relationship between the curvature of the radio spectrum of a galaxy and the degree of CR proton calorimetry it achieves, a prediction that will be testable by upcoming $\gamma$-ray observations.

\subsection{Why is the FRC straight and narrow -- a conspiracy?}
\label{ssec:FRCexplanation}

We have seen that our models reproduce a straight, narrow FRC that agrees well with observations. Our goal in this section is to understand \textit{why}, and to investigate whether the FRC is indeed the result of a conspiracy of cancellations.

\subsubsection{From FIR luminosity to SFR}

\begin{figure}
	\includegraphics[width=\columnwidth]{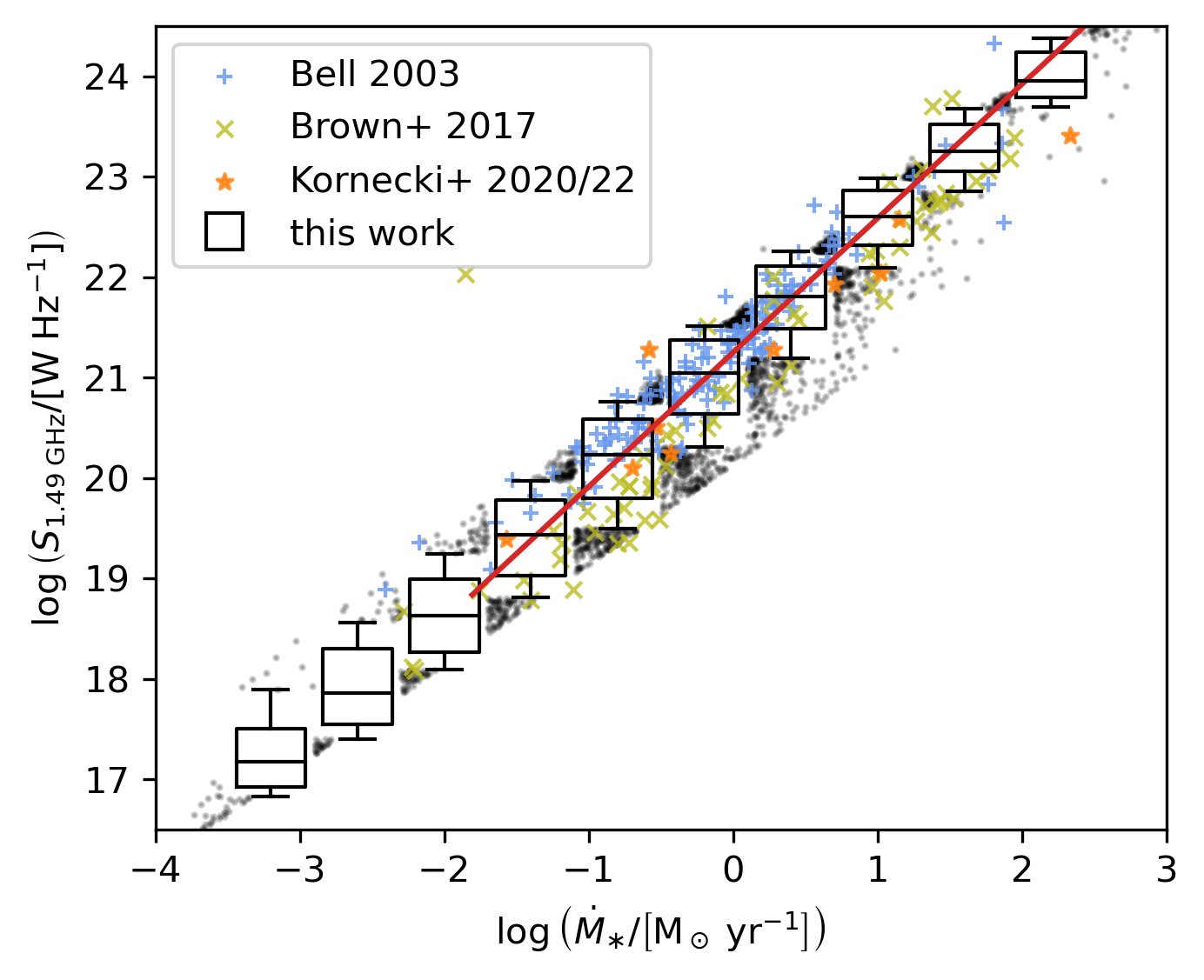}
    \caption{1.49 GHz flux density versus star formation rate as predicted by our model for CANDELS galaxies and for observed samples. All symbols have the same meaning as in \autoref{fig:FIRradio}, except that we have replaced FIR luminosity on the horizontal axis with SFR. The red line shows the fit between the indicated limits, given in \autoref{eq:SFRradio} for which $S_\nu \propto \dot{M_*}^{1.3}$ (at $\nu$ = 1.49 GHz).}
    \label{fig:SFRradio}
\end{figure}

As a first step toward this, we remove the influence of the SFR to FIR scaling. In \autoref{fig:SFRradio} we therefore show the SFR-radio correlation; the data we show here are identical to those shown in
\autoref{fig:FIRradio}, 
with the exception that we have not used \autoref{eq:LFIR} to convert the SFRs from CANDELS galaxies (which are derived by fitting their rest frame optical-UV fluxes to stellar population models) to IR luminosities. For the observational data, we utilise the reported SFRs of \citet{2020A&A...641A.147K, 2022A&A...657A..49K} and \citet{2017ApJ...847..136B}, and invert the relation from \autoref{eq:LFIR} to convert the FIR luminosities in \citet{2003ApJ...586..794B} to SFR.
Fitting a power law to our predicted distribution for the CANDELS sample yields
\begin{equation}
\log\left(\frac{S_{\rm 1.49 GHz}}{\rm W \,Hz^{-1}} \right) = 1.3 \log\left(\frac{ \dot{M}_{\ast}}{ \rm M_{\odot} \, yr^{-1}} \right) + 21.3. 
\label{eq:SFRradio}
\end{equation}
This relation shows a similar dispersion as the FRC shown in 
\autoref{fig:FIRradio}
but a slightly steeper slope $\approx 1.3$, compared to 1.2 for the FRC. 
This increase stems from the SFR to FIR conversion, which scales $\propto \dot{M}_{\ast}^{1.1}$ (\autoref{eq:LFIR}) in our model. The difference is consistent with the hypothesis of \citet{2010ApJ...717....1L} that the reduced UV opacity of low SFR systems, which gives rise to the super-linear SFR-FIR conversion because low SFR systems reprocess less of their starlight into the IR, contributes to flattening the FRC. However, this effect is also clearly non-dominant, since it changes the slope only by $0.1$.\footnote{Flattening of the FRC via the non-linearity of the SFR-FIR conversion may be a much stronger effect for very dim systems, e.g., see \citet{2003ApJ...586..794B}. However our galaxy sample -- and the sample of DustPedia galaxies from which we derive our SFR-IR conversion (see \citetalias{roth+2023}) -- contains very few galaxies at the extreme low end of SFRs. Thus the effect is not dominant for our sample.}

\subsubsection{Secondaries and synchrotron calorimetry}
\label{sssec:secondaries}

Now that we have eliminated the FIR-SFR conversion as a significant contributor to the FRC, we are left with two related questions: (1) why is the slope in the SFR-radio correlation 1.3 instead of 1.0 (as one might naively guess), and (2) why are there no visible bends or locations where the scatter of the correlation becomes large (as one might expect if secondary electrons or non-synchrotron losses were to vary greatly with SFR)? To answer these two questions, we can envision the SFR-radio correlation as being determined by two factors: the fraction of SFR-supplied power that is eventually channeled into CR electrons (which is potentially altered by the contribution of secondary electrons), and the 
fraction of that power that emerges as observable synchrotron emission.

\paragraph{The secondary electron power supply}

\begin{figure}
	\includegraphics[width=\columnwidth]{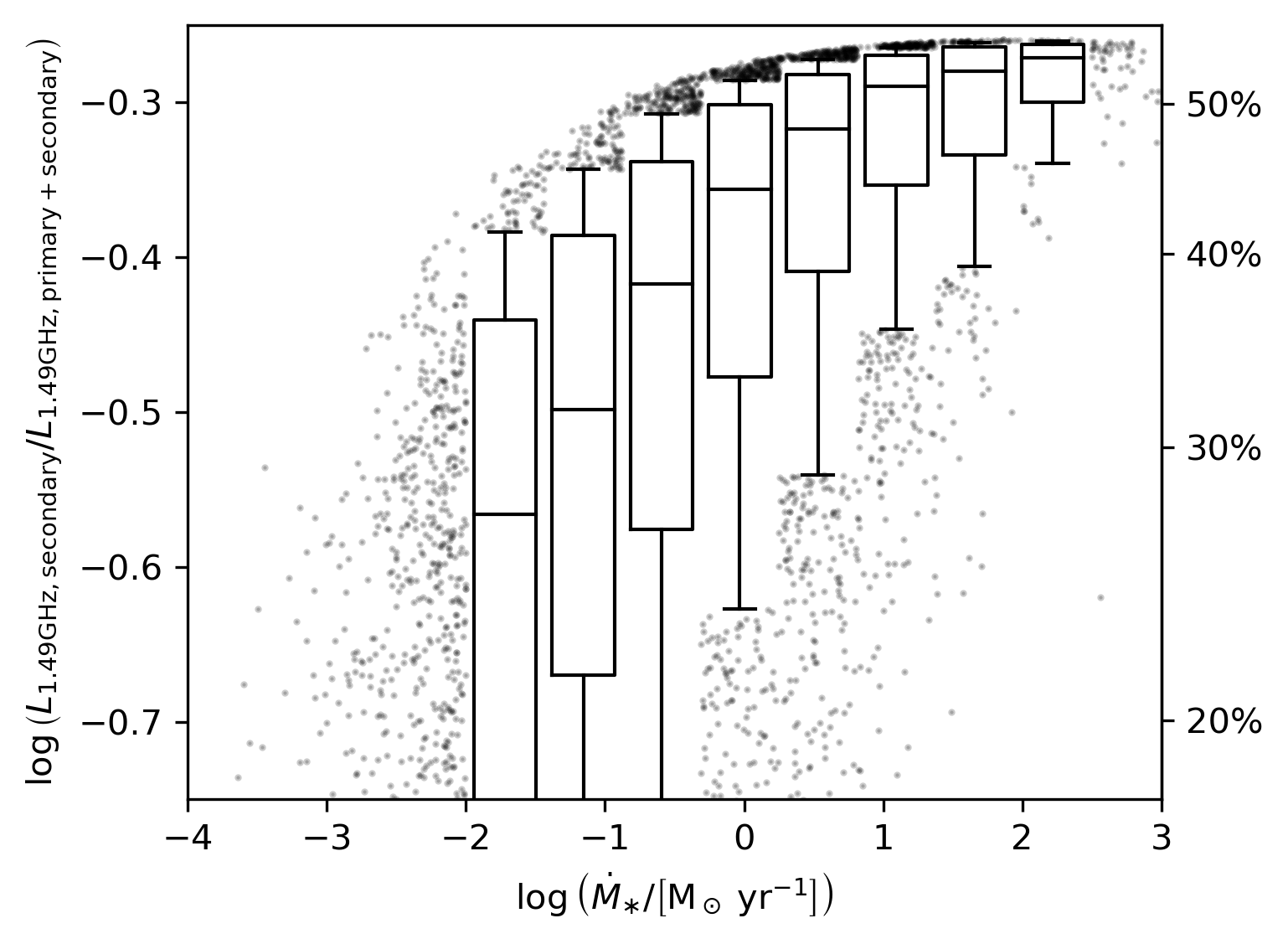}
    \caption{The ratio of secondary to total CR electron contribution to radio emission at 1.49 GHz as a function of star formation rate, as predicted by our model applied to the CANDELS sample. Boxes, whiskers, and black points have the same meaning as in \autoref{fig:FIRradio}. The right vertical axis simply shows the value on the left axis expressed a percentage.}
    \label{fig:Lsec_Ltot}
\end{figure}

To quantify the first of these factors, in \autoref{fig:Lsec_Ltot} we show the ratio of secondary-driven to total (primary plus secondary) radio emission for the galaxies in our sample; recall that \textsc{CONGRuENTS} separates these contributions, allowing us to calculate them independently. The figure shows that, with increasing star-formation rate, the contribution of secondary CR electrons to the total emission at 1.49 GHz becomes increasingly important, and indeed dominant over the primary contribution in systems towards the extreme end of the star formation rate distribution. The source of these additional secondary electrons and positrons is the decay of charged pions produced in the collisions of CR protons with the ISM. Increasing proton calorimetry in denser, high SFR systems thus leads to an increasing contribution from secondaries to the total emission that eventually dominates over primaries. However, we also see from the figure that this changeover induces a 
{\it prima facie}
surprisingly 
mild
change in the CR electron power supply.

We can understand this from an energetics point of view. In our model, 10\% of supernova energy is injected into CR protons and a further 2\% in primary CR electrons and, with these numbers, we would not expect equipartition between primary and secondary electrons to be 
violated
by much. For full proton calorimetry $\approx 1/9$th of primary CR proton energy is deposited in secondary electrons -- $\approx 2/3$ of proton energy goes into charged pions, of which $\approx 1/2$ produces muons, of which $\approx 1/3$ gets deposited in an electron or positron. 
This approximately corresponds to primary-secondary equipartition at the critical energy (within say a factor of two), albeit modulated by the somewhat different spectral shapes of the primary and secondary electron distributions (considering that the former are not subject to the pion production threshold energy). This spectral effect leads to the observed slight further enhancement of the secondary excess at high SFRs that is not explained from the purely energetic argument. %
Nonetheless, the overall conclusion to be drawn from this discussion, and from \autoref{fig:Lsec_Ltot}, is that the total contribution of secondary electrons to radio emission, while important, increases by at most a factor of two over a $\approx 5$ decade range in SFR, corresponding to a slope of $\approx 0.05-0.1$ in a log-log plot.

We can also use this analysis to comment on the dependence of our results on our assumed 5:1 ratio of SN energy deposited in primary protons versus primary electrons; while this ratio is well-motivated by both direct and indirect evidence, as described in \autoref{sec:methods}, factor of $\sim 2$ variations are clearly plausible given any realistic assessment of the uncertainties. If the proton-to-electron ratio were a factor of $\sim 2$ larger, our analysis suggests that transition from primary electrons dominating the synchrotron emission to secondaries dominating would occur at a star formation rate $\sim 0.5$ dex lower than that shown in \autoref{fig:Lsec_Ltot}, and the increase in synchrotron emission due to secondaries as we go from the most weakly to the most strongly star-forming galaxies would be a factor of $\approx 4$ rather than $\approx 2$ as in our fiducial case. Correspondingly, the contribution of secondaries to the slope would of the FIR-radio relation would rise to $\approx 0.1-0.15$. Conversely, if we were to posit at 5:2 rather than a 5:1 ratio of primary protons to primary electrons, the maximum secondary electron contribution to synchrotron radiation would drop from half to $\approx 1/4$, so secondaries would never be significant and an increasing secondary contribution would not affect the slope of the FIR-radio correlation at all. In summary, our fiducial choice of proton-to-electron ratio means that secondaries steepen the FIR-radio correlation by $\approx 0.05-0.1$ dex, and a plausible estimate of the uncertainty suggests that this could be as small as $\approx 0$ and as large as $\approx 0.1-0.15$ dex. Given the observational uncertainties in the slope, these variations are not particularly significant.

\paragraph{The synchrotron calorimetry fraction}
\label{sec:haloICdiscBS}

\begin{figure}
	\includegraphics[width=\columnwidth]{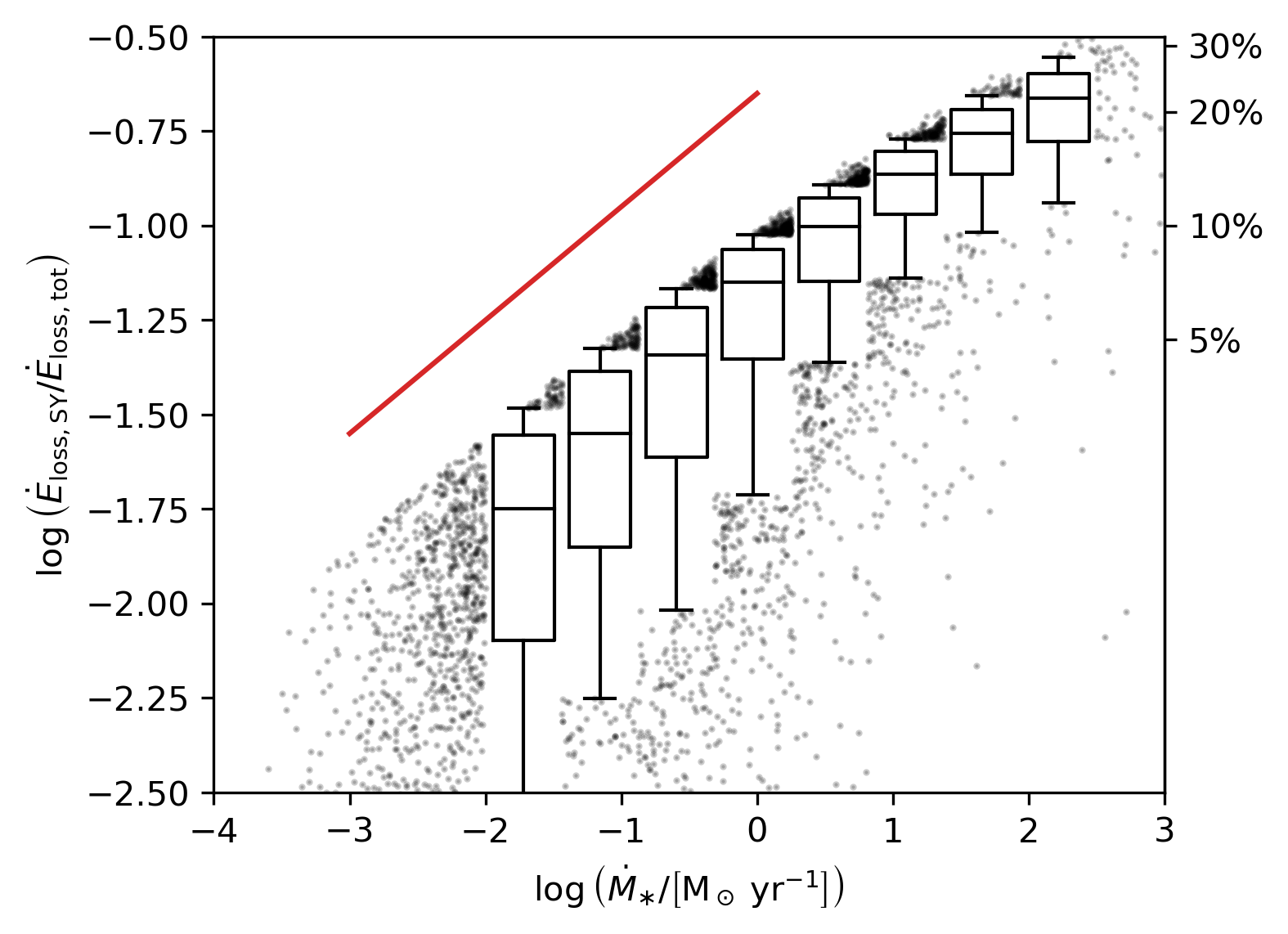}
    \caption{The logarithm of the ratio of the loss rate into synchrotron emission to the total loss rate into all loss channels for electrons in the galactic disc whose energies are such that the critical frequency for synchrotron emission is 1.49 GHz. Boxes, whiskers, and points have the same meaning as in \autoref{fig:FIRradio}, and the right vertical axis shows the same values as on the left vertical axis, just converted to percentages. The red line shows a power law relation $\propto \dot{M}_{\ast}^{0.3}$ for reference. The total loss rate here includes losses into synchrotron emission, bremsstrahlung, inverse Compton emission, ionisation, and diffusion into the halo.}
    \label{fig:ElossSy_Elosstot}
\end{figure}

The second factor that influences the SFR-radio correlation is the fraction of CR electron power that is eventually radiated as synchrotron emission and observed at 1.49 GHz. To quantify this factor, in \autoref{fig:ElossSy_Elosstot} we plot the ratio of the synchrotron loss rate to the total loss rate, computed for electrons in the disc whose energy $E_{\rm e}$[\footnote{We note that $E_{\rm e}$ will gain a dependence on redshift if considering the critical frequency in the observer rather than the emitter frame as we do here.}] is such that their synchrotron critical frequency is $\nu_{\rm c} = E_{\rm e}^2 e B / 2 \pi m_{\rm e}^3 c^5 = 1.49$ GHz; here $B$ is the magnetic field strength in the disc and we note that $\nu_{\rm c}$ samples increasingly low CR energies for increasing magnetic field strength. Before interpreting this plot, we pause here to point out two important methodological considerations in its construction. First, we are focusing on the electron energy range that dominates emission at 1.49 GHz, since that is what is generally observed. However, since the synchrotron critical frequency depends on the magnetic field strength and this varies from galaxy to galaxy, this choice means that the quantity shown in \autoref{fig:ElossSy_Elosstot} is not evaluated at the same CR electron energy in every galaxy, a point that will become important below. Second, we focus on losses in the disc rather than the halo because the disc in general contributes much more synchrotron luminosity. This is partly because in higher SFR galaxies only a relatively small fraction of CR electrons escape into the halo at all, and partly because inverse Compton, rather than synchrotron, losses dominate in galaxy halos due to the fact that, for our adopted plane-parallel geometry (and as expected in real galaxies), magnetic energy densities (which for reasons of hydrodynamic stability cannot be too far out of equipartition with gas thermal plus kinetic energies) fall off away from galactic discs much more rapidly than photon energy densities.

\autoref{fig:ElossSy_Elosstot} allows a few immediate conclusions. First, our models are not synchrotron-calorimetric, since only a sub-dominant fraction of CR energy gets radiated in synchrotron emission; this fraction never exceeds $\approx 30\%$ and is of the order of only a few percent for dim systems. Second, however, the fraction of energy radiated into synchrotron varies only weakly with SFR, and does so in a roughly power law fashion, i.e., the behaviour shown in \autoref{fig:ElossSy_Elosstot} can be described reasonably well by a straight line. Quantitatively, the synchrotron calorimetry fraction increases from $\approx 2\%$ to $\approx 30\%$ over five decades on star formation rate, corresponding to a slope of $\approx 0.25$ in log-log. This, combined with the slope of $\approx 0.05-0.1$ describing the increase in secondary electron contribution with SFR, is sufficient to explain the observed slope of $\approx 1.3$ in the SFR-radio correlation, i.e., we can attribute the mild super-linearity of 0.3 in the slope to a contribution of $\approx 0.25$ from the gradual increase in synchrotron calorimetry fraction with SFR, and $\approx 0.05-0.1$ from the mild increase in total CR electron power due to the contribution from secondaries.

\paragraph{The heart of the conspiracy: why is the synchrotron calorimetry fraction a nearly 
featureless
power law?}

We have now almost 
penetrated to
the centre of the conspiracy that gives rise to the straight, narrow FRC, which we have seen arises from the combination of nearly pure power law behaviour in both the secondary electron contribution and the degree of synchrotron calorimetry. While the former is easy to understand on energetic grounds, as explained above, the latter is somewhat more mysterious: why does the fraction of CR electron power radiated into 1.49 GHz synchrotron emission change by only a factor of $\sim 10$, and nearly linearly (in log-log), as the SFR increases by a factor of $\sim 10^5$?

\begin{figure}
	\includegraphics[width=\columnwidth]{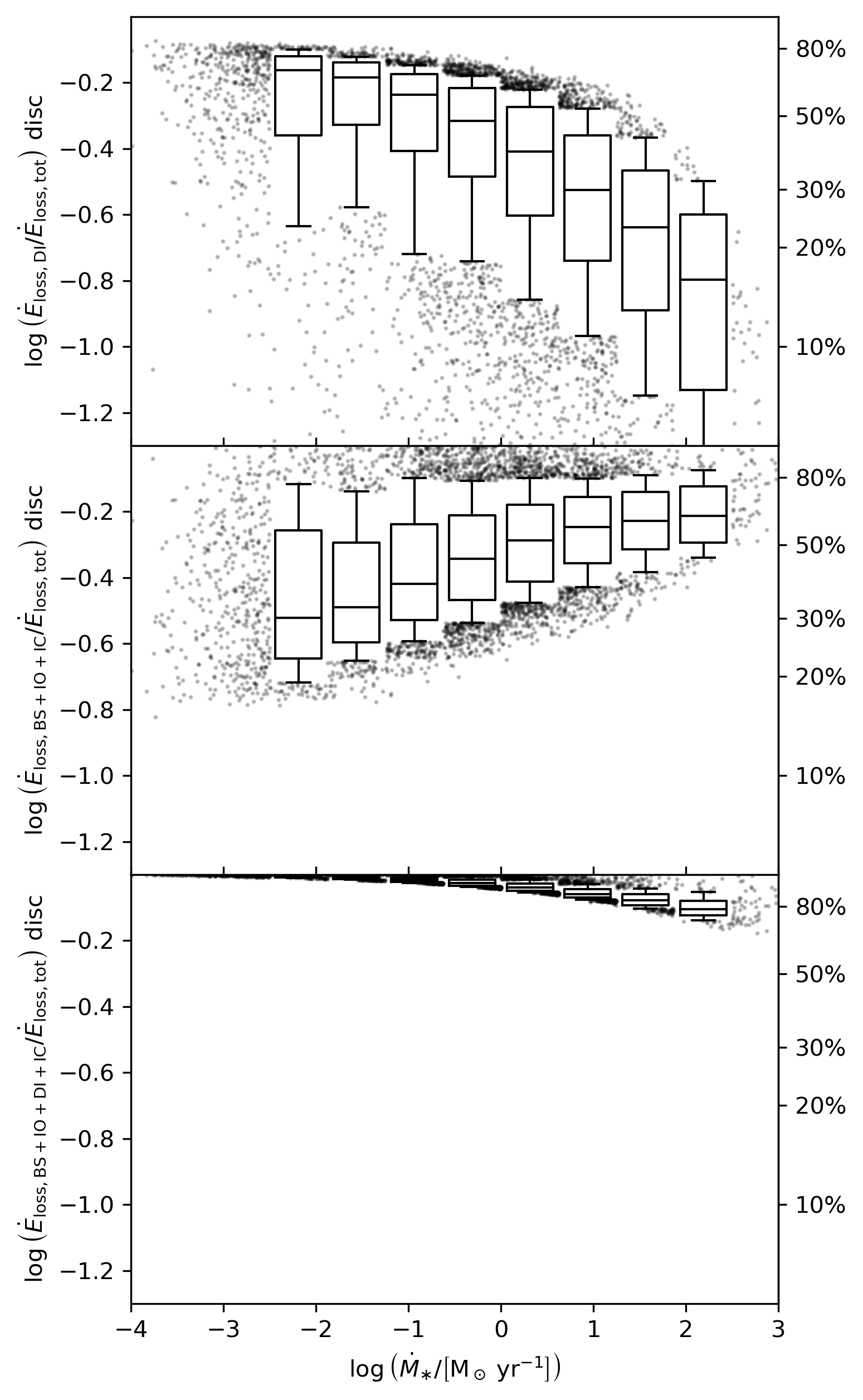}
    \caption{From top to bottom we show on logarithmic scale: (top) the ratio of the diffusive loss rate from the disc into the halo over the total loss rate;
    (middle) the combined loss rate into bremsstrahlung, ionisation and inverse Compton over the total loss rate; and (bottom) the combined loss rates for all four loss processes (i.e., all but synchrotron emission) over the total loss rate. 
    All plots are for electrons in the disc with critical synchrotron frequency of 1.49 GHz.
    Boxes, whiskers and black points have the same meaning as in \autoref{fig:FIRradio}. On the right frame edges we show  corresponding percentages.}
    \label{fig:Eloss3_Elosstot}
\end{figure}

To answer this question it is helpful to consider the loss processes that compete with synchrotron, and thus represent alternative sinks for CR electron power in the disc: bremsstrahlung, ionisation, inverse Compton, and diffusion into the halo. We show the fractional contributions of these processes, again evaluating at the electron $E_{\rm e}$ for which the critical frequency is 1.49 GHz, in \autoref{fig:Eloss3_Elosstot}; in this figure, we have separated out losses from diffusive escape into the halo (top panel), all other non-synchrotron losses in the disc (middle panel), and the sum of all non-synchrotron loss processes (bottom panel). As the figure makes clear, diffusion into the halo is the dominant loss mechanism for CR electrons in galaxies with SFRs $\lesssim 1$ M$_\odot$ yr$^{-1}$, but becomes sub-dominant at higher SFRs. By contrast, bremsstrahlung, ionisation, and inverse Compton losses show exactly the opposite pattern, remaining subdominant at low SFRs but dominating at SFRs $\gtrsim 1$ M$_\odot$ yr$^{-1}$. The net effect of this crossover, however, is that the \textit{sum} of the non-synchrotron loss processes always dominates over synchrotron, and the amount by which they dominate varies only weakly with SFR across the full range in SFR. This leaves the fraction of the losses that go into 
the residual -- i.e.,
synchrotron emission -- (c.f.~\autoref{fig:ElossSy_Elosstot}) with only a weak SFR dependence.

While this might at first appear to indeed be a conspiracy or an accident, this appearance is deceptive. Rather it is a natural and understandable consequence of (1) the way that magnetic field strength interacts with observations at a fixed frequency to select electrons of differing energies in differing galaxies; (2) the turbulent dynamo, which constrains the strength of galactic magnetic fields; (3) hydrostatic balance, which constrains galactic gas densities; and (4) the tight scaling of various galaxy properties with $\dot{M}_\ast$ along the SFMS. We can use these principles and relations to see directly why the synchrotron calorimetry fraction should scale only weakly with star formation rate.

First consider how the loss times for synchrotron emission and for the dominant loss processes -- diffusion at low SFR, ionisation and bremsstrahlung at high SFR -- scale with CR electron energy $E_{\rm e}$, gas density $n_\mathrm{H}$, and magnetic field strength $B$; these scalings are
\begin{equation}
    \tau_\mathrm{SY} \propto B^{-2} E_{\rm e}^{-1}
    \quad
    \tau_\mathrm{BS} \propto n_\mathrm{H}^{-1}
    \quad
    \tau_\mathrm{IO} \propto n_\mathrm{H}^{-1} E_{\rm e}
    \quad
    \tau_\mathrm{DI} \propto h_g/v_\mathrm{st},
\end{equation}
where $h_g$ is the gas scale height, $v_\mathrm{st}$ is the characteristic CR streaming speed, and the subscripts SY, BS, IO, and DI indicate synchrotron, bremsstrahlung, ionisation, and diffusion, respectively. While inverse Compton losses are important in the halo, and at higher energy in the disc, they are never dominant for the low-energy CRs between $\sim$0.1 to $\sim$10 GeV that dominate 1.49 GHz synchrotron emission. Since we are observing at a fixed frequency, however, the characteristic electron energy to which we are sensitive is itself a function of the magnetic field strength, as $E_{\rm e} \propto 1/\sqrt{B}$, and these scalings therefore reduce to
\begin{equation}
    \tau_\mathrm{SY} \propto B^{-3/2}
    \quad
    \tau_\mathrm{BS} \propto n_\mathrm{H}^{-1}
    \quad
    \tau_\mathrm{IO} \propto n_\mathrm{H}^{-1} B^{-1/2}
    \quad
    \tau_\mathrm{DI} \propto h_g/v_\mathrm{st}.
\end{equation}

Next we make use of two important physical principles that are embedded in our model: hydrostatic balance and turbulent dynamo action. The dynamo both forces the streaming speed $v_\mathrm{st}$ to scale with the gas velocity dispersion $\sigma_\mathrm{g}$ (since the dynamo saturates at roughly fixed Alfv\'en Mach number) and enforces rough equipartition between turbulent and magnetic pressures, requiring $B\propto \sigma_\mathrm{g} \sqrt{n_\mathrm{H}}$. Hydrostatic balance is not reducible to quite such a simple power law scaling, since it depends on the combination of gas and stellar gravity, but as a rough approximation for the purposes of this argument we can take it to require $\sigma_\mathrm{g}/h_\mathrm{g} \propto \sqrt{n_\mathrm{H}}$ (c.f.~Equation 3 of \citetalias{roth+2023}), a scaling that becomes exact when gas self-gravity dominates over stellar gravity. Inserting these expressions for $B$, $\sigma_\mathrm{g}$, and $h_\mathrm{g}$ into our estimates of the loss times gives us
\begin{equation}
    \tau_\mathrm{SY} \propto \sigma_\mathrm{g}^{-3/2} n_\mathrm{H}^{-3/4}
    \quad
    \tau_\mathrm{BS} \propto n_\mathrm{H}^{-1}
    \quad
    \tau_\mathrm{IO} \propto \sigma_\mathrm{g}^{-1/2} n_\mathrm{H}^{-5/4}
    \quad
    \tau_\mathrm{DI} \propto n_\mathrm{H}^{-1/2}.
\end{equation}

Next we invoke two empirical scaling relations that are observed to exist between galaxies' internal properties -- $\sigma_\mathrm{g}$ and $n_\mathrm{H}$ -- and their bulk properties, $\dot{M}_\ast$, $M_\ast$ and $R_e$; again, these relations are embedded in our model. The first of these is the relationship between $\sigma_\mathrm{g}$ and $\dot{M}_\ast$ whereby more rapidly star-forming galaxies have higher velocity dispersions (\citealt{2019MNRAS.486.4463Y}; \citetalias{roth+2023}), and the second is the so-called
extended Kennicutt-Schmidt law \citep{2011ApJ...733...87S}, which links galaxies' gas surface densities to their star formation and stellar surface densities -- this relationship coupled with hydrostatic balance in turn constrains $n_\mathrm{H}$. Using these relationships (which we refrain from writing down explicitly for brevity) in the scalings above yields
\begin{eqnarray}
    \tau_\mathrm{SY} \propto R_e^{2.25} M_\ast \dot{M}_\ast^{-1.69}
    & \quad &
    \tau_\mathrm{BS} \propto R_e^3 M_\ast \dot{M}_\ast^{-1.6}
    \nonumber \\
    \tau_\mathrm{IO} \propto R_e^{3.75} M_\ast \dot{M}_\ast^{-1.91}
    & \quad &
    \tau_\mathrm{DI} \propto R_e^{1.5} M_\ast^{0.5} \dot{M}_\ast^{-0.8}.
\end{eqnarray}
Although we have already eliminated the magnetic field strength from our equations, we can use the same method to yield a relation of $B \simeq 37 \ (R_{\rm e}/\rm kpc)^{-1.5} ( \dot{M}_{\ast}/\rm M_{\odot} yr^{-1} ) (M_{\ast}/\rm 10^{9} \ M_{\odot})^{-0.5} \ {\rm \mu G}$.
Finally, we must consider the fact that $\dot{M}_\ast$ and $M_\ast$ are themselves correlated along the SFMS -- indeed, this correlation is what \textit{defines} the SFMS \citep{2014ApJS..214...15S}. For our sample, this correlation is roughly $M_\ast \propto \dot{M}_\ast^{0.74}$, and using this relationship in the expression above to eliminate $\dot{M}_*$ we have
\begin{eqnarray}
    \tau_\mathrm{SY} \propto R_e^{2.25} \dot{M}_\ast^{-0.94}
    & \quad &
    \tau_\mathrm{BS} \propto R_e^3 \dot{M}_\ast^{-0.86}
    \nonumber \\
    \tau_\mathrm{IO} \propto R_e^{3.75} \dot{M}_\ast^{-1.17}
    & \quad &
    \tau_\mathrm{DI} \propto R_e^{1.5}  \dot{M}_\ast^{-0.43}.
\end{eqnarray}

It is straightforward to see how these scalings generate the correlation we observed between synchrotron calorimetry and star formation rate. Since $\tau_\mathrm{DI}$ and $\tau_\mathrm{IO}$ have the weakest and strongest dependence on $\dot{M}_\ast$, respectively, diffusive losses should dominate at low SFRs and ionisation ones at high SFR, with bremsstrahlung forming a bridge in between, exactly as we observe. This in turn means that in the low SFR regime the fraction of power radiated as synchrotron emission should scale fairly weakly with SFR, as $\tau_\mathrm{DI}/\tau_\mathrm{SY} \propto \dot{M}_\ast^{0.51}$, in the mid-SFR regime as $\tau_\mathrm{BS}/\tau_\mathrm{SY} \propto \dot{M}_\ast^{0.08}$, and in the high SFR regime as $\tau_\mathrm{IO}/\tau_\mathrm{SY} \propto \dot{M}_\ast^{-0.23}$. Of course this analysis omits the scaling between $\dot{M}_*$ and $R_e$, and relies on an oversimplification of our treatment of hydrostatic balance, as noted above. Nonetheless, this analysis establishes the basic point that we should expect the synchrotron calorimetry fraction to show only a weak dependence on star formation rate, one that, given the large scatter induced by varying $R_e$ at fixed $\dot{M}_\ast$ (and by the fact that the SFMS is not infinitely tight) can easily appear as a single power law with a shallow slope.

Most importantly, this analysis shows that this behaviour is \textit{not} simply a conspiracy or an accident. It is a consequence of observing at a fixed frequency (implying that we observe lower energy CRs in more highly magnetised galaxies; \citealt{2010ApJ...717....1L}), basic considerations of energy and pressure balance (hydrostatic equilibrium and turbulent dynamo) that dictate the balance between synchrotron and other loss processes as a function of ISM properties, and the tight empirical relations between galaxies' star formation rates and those internal properties that appear to be imposed by the process of galaxy formation.

\subsection{The origin of the radio spectral index distribution of star forming galaxies}
\label{ssec:specindex}

The radio spectral index is readily measured observationally and can provide insights into the loss mechanisms that shape the cosmic-ray spectrum; indeed, recall that the spectral index being $\approx 0.6-0.8$ rather than $\approx 1.1$ is perhaps the strongest piece of evidence that the CR electrons in galaxies cannot lose most of their energy in synchrotron emission (nor, for that matter, in IC). 
The fact that our models correctly reproduce the observed spectral index (c.f.~\autoref{fig:alphaSFR}) provides strong confirmation that the balance between the various loss processes in our model is about right, and offers the opportunity for us to explain the origin of the spectral index using the same approach with which we explained the FRC in \autoref{ssec:FRCexplanation}.

To first order, the non-thermal radio spectral index of a source depends on the shape of the injection spectrum at the critical energy corresponding to the radio observation frequency, and the energy dependence of the loss processes responsible for setting the CR electron steady state spectrum at that energy. 
For an injection spectrum in the form of a power law in energy $d^2N/dEdt \propto E^{-q}$, and a cooling process for which the energy loss rate as a function of energy varies as $\dot{E} \propto E^{\gamma_{\rm cool}}$, the radio spectral index $S_{\rm \nu} \propto \nu^{-\alpha}$ can be approximated by integrating a simplified kinetic equation in which we neglect the catastrophic nature of inverse Compton and bremsstrahlung interactions, and treat them as continuous instead; while \textsc{CONGRuENTS} solves the full kinetic equation numerically (see Equation 21 in \citetalias{roth+2023}), the simplified version is useful to gain insight because  we can solve it analytically in two limiting cases: i) the thick target (obtained by setting the escape term to zero) and ii) the thin target (obtained by setting the loss term to zero).
For the former,
the steady state spectral index becomes $\alpha = (q + \gamma_{\rm cool})/2 - 1$, while for the latter we instead obtain $\alpha = (q - 1)/2$ as long as the escape fraction is independent of CR electron energy (as is the case in our model for CR electrons at the energies of interest for synchrotron production). 
Synchrotron and inverse Compton losses both have $\gamma_{\rm cool} = 2$, bremsstrahlung has $\gamma_\mathrm{cool} = 1$, and ionisation losses have $\gamma_{\rm cool} = 0$.

Thus for a primary electron injection index of $q = 2.2$, our fiducial choice, and in the limit of a thick target informed by a single, dominant loss process, we obtain a characteristic synchrotron spectral index of $\alpha = 1.1$ for the case that synchrotron itself or inverse Compton is this dominant loss process, $\alpha=0.6$ if bremsstrahlung dominates, and $\alpha = 0.1$ for ionisation losses dominant. If galaxies are mostly thin, on the other hand and thus diffusion dominated, we expect $\alpha = 0.6$.
Finally, free-free emission has, independently of these considerations, a characteristic radio index $\alpha \simeq 0.1$.

As shown in \autoref{fig:Eloss3_Elosstot}, we do not find that a single mechanism dominates for all galaxies at all star formation rates. Instead, galactic discs tend to present thin targets for CR electrons at low star formation rates, with bremsstrahlung taking over at middle star formation rates, and ionisation becoming important only at the very highest star formation rates. However, at any given star formation rate there is considerable scatter, driven by the scatter in galactic effective radii and stellar masses. Thus several different loss processes combine to inform the radio spectral index. This is broadly consistent with the findings of \citet{2006ApJ...645..186T} and \citet{2010ApJ...717....1L}.

\begin{figure*}
	\includegraphics[width=\textwidth]{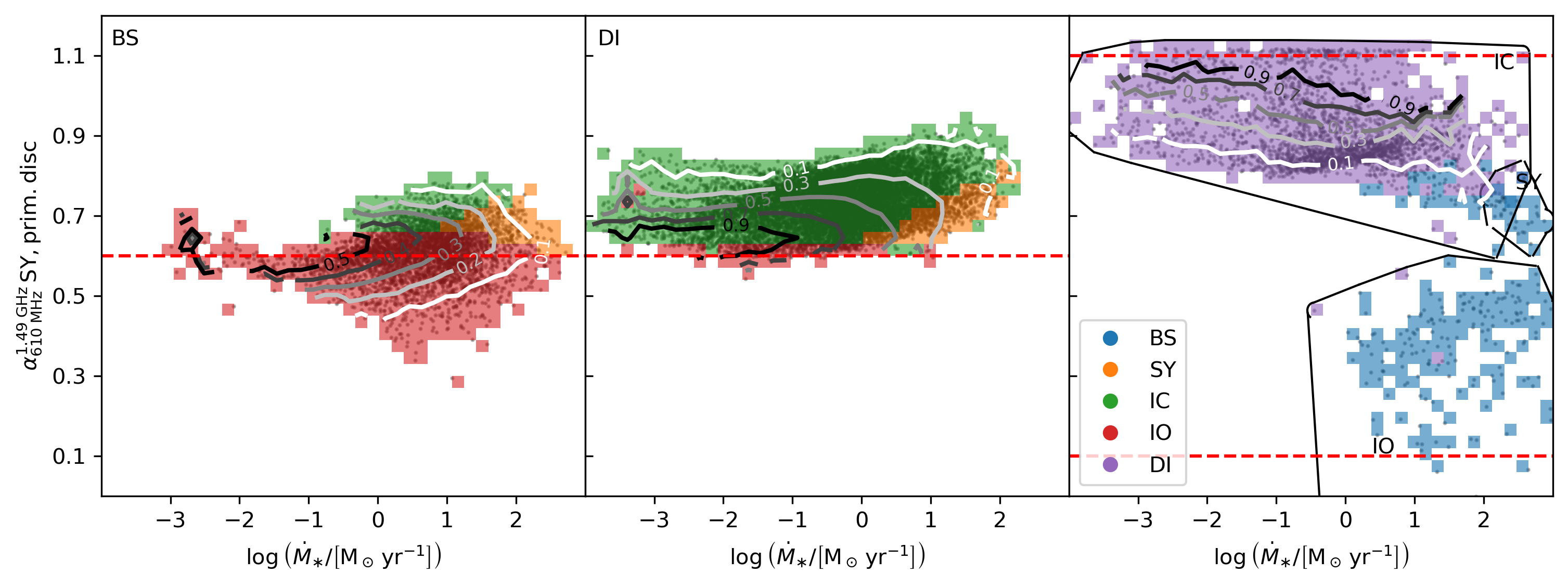}
    \caption{The synchrotron spectral index from 610 MHz to 1.49 GHz for emission by disc primary electrons vs.~star formation rate.  Points indicate individual galaxies, and galaxies are divided into different panels according 
    to which processed (as labelled) is the dominant loss process for electrons whose critical frequency is 1.49 GHz.
    Loss and escape processes are bremsstrahlung (BS), (energy-independent) diffusion from disc into the halo (DI), inverse Compton (IC), synchrotron (SY) and ionisation losses (IO); the latter three are combined in panel three. 
    The colour coding specifies the next most important loss process that has a different energy dependence from the primary loss process, with the mapping from colour to process as indicated in the legend. Contour lines show the logarithm of the ratio of the energy loss rate of the primary to the next most important loss process. The red dashed lines show the spectral index we would expect for a population whose losses are fully dominated by the primary process of each panel, where in the last panel 1.1 corresponds to inverse Compton and synchrotron and 0.1 to ionisation.}
    \label{fig:alpha_lossbins}
\end{figure*}

We can disentangle these influences and see how they combine to produce the observed spectral index distribution with the aid of \autoref{fig:alpha_lossbins}. In this figure we plot the spectral index for 1.49 GHz synchrotron emission for primary electrons in the disc as a function of star formation rate; for simplicity in this plot we omit the effects of free-free opacity, and we focus on the disc rather than the halo and on primaries rather than secondaries because, for most galaxies, these are dominant. The key feature of \autoref{fig:alpha_lossbins} is that, rather than showing all galaxies together, we separate galaxies into three panels according to which loss process (as labelled) is dominant at the critical energy;
the first two panels show the most common cases of bremsstrahlung- and diffusion-dominated galaxies, while the last panel consolidates the rarer cases of inverse Compton-, synchrotron- and ionisation-dominated systems. Examining the figure, we see a clear dichotomy between the diffusion- and bremsstrahlung-dominated galaxies, which sit near but slightly above the value $\alpha_\mathrm{610\,MHz}^{1.49\,\mathrm{GHz}} = 0.6$ we would expect if those processes were completely dominant, the ionisation-dominated galaxies which sit at lower $\alpha_{1.49\,\mathrm{GHz}}$, and the inverse Compton- and synchrotron-dominated galaxies, which sit at higher $\alpha_\mathrm{610\,MHz}^{1.49\,\mathrm{GHz}}$.

In addition to showing the dominant loss process, \autoref{fig:alpha_lossbins} also indicates the most important 
\textit{sub-dominant}
loss mechanism for which a different spectral index would be expected. To be precise, we divide the $(\dot{M}_\ast, \alpha_\mathrm{610\,MHz}^{1.49\,\mathrm{GHz}})$ plane into cells, and within each cell containing at least one galaxy we identify the {\it next most important loss mechanism with a different $\gamma_{\rm cool}$} for the majority of galaxies in that cell. Thus for example if we examine the left panel of \autoref{fig:alpha_lossbins}, which shows galaxies for which bremsstrahlung ($\gamma_\mathrm{cool} = 1$, $\alpha_\mathrm{610\,MHz}^\mathrm{1.49\,GHz} = 0.6$) is the dominant cooling process, red cells indicate galaxies for which ionisation ($\gamma_\mathrm{cool} = 0$, $\alpha_\mathrm{610\,MHz}^\mathrm{1.49\, GHz} = 0.1$) is the second-most important cooling process with different $\gamma_\mathrm{cool}$, orange cells indicate synchrotron ($\gamma_\mathrm{cool} = 2$, $\alpha_\mathrm{610\,MHz}^\mathrm{1.49\, GHz} = 1.1$) as the next-most important process, and green cells indicate inverse Compton (also $\gamma_\mathrm{cool} = 2$, $\alpha_\mathrm{610\,MHz}^\mathrm{1.49\, GHz} = 1.1$) as next-most important. On top of these colours, we also show contours that indicate the log ratio of the loss rates from the indicated primary and secondary loss processes, and thus quantify how dominant the former actually is.

From the colours, it is clear that the action of the next-most-important loss process is to push the spectral index away from the characteristic index expected from the primary loss process, and toward the value favoured by the secondary one. Thus for example in the central panel, diffusion is the dominant process, which would favour $\alpha_\mathrm{610\,MHz}^\mathrm{1.49\, GHz} = 0.6$, but for many of these galaxies inverse Compton is the next most important process with different $\alpha_\mathrm{cool}$ (green region), and for these galaxies IC effects drag the spectral index upward some distance toward the value $\alpha_\mathrm{610\,MHz}^\mathrm{1.49\, GHz} = 1.1$ that it favours, with the upward displacement largest for those galaxies where inverse Compton is only slightly sub-dominant (primary to secondary ratio near unity). Similarly in the rightmost panel we see a small sample of galaxies for which inverse Compton dominates and $\alpha_\mathrm{610\,MHz}^\mathrm{1.49\, GHz}$ is $\sim 1$ (upper box), but for these galaxies diffusion is the next most important process, which pulls their spectral indices downward somewhat. 

\begin{figure}
	\includegraphics[width=\columnwidth]{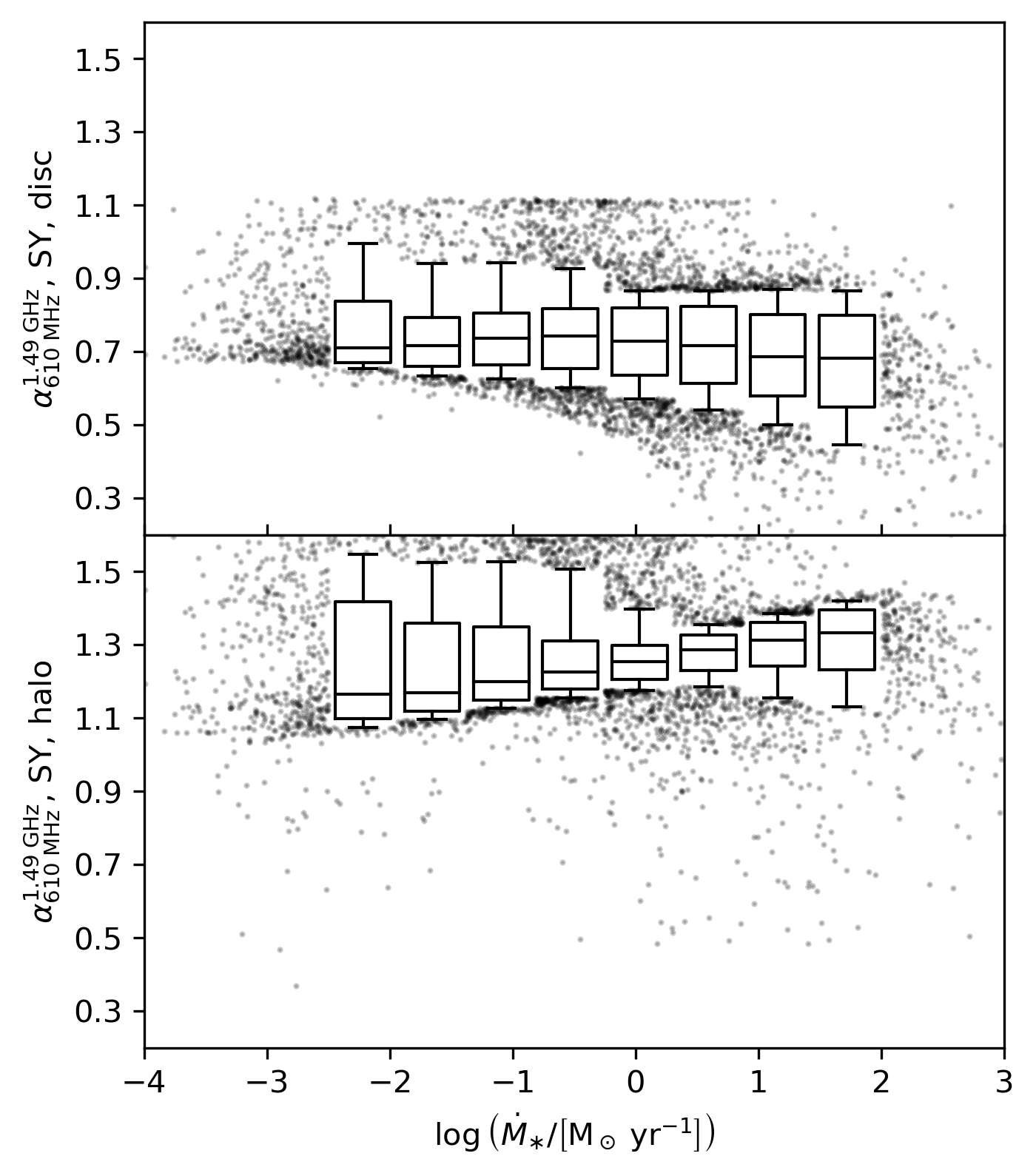}
    \caption{The synchrotron spectral index between 610 MHz and 1.49 GHz for emission by primary and secondary CR electrons in the disc (top panel) and the halo (bottom pane) vs.~star formation rate. Boxes, whiskers and black points have the same meaning as in \autoref{fig:FIRradio}.}
    \label{fig:alpha_dischalo}
\end{figure}

We have therefore identified the primary reason why $\alpha_\mathrm{610\,MHz}^\mathrm{1.49\, GHz}$ sits in the range $\approx 0.6-0.8$ for most galaxies: it is because for most galaxies the dominant loss process for primary electrons in the disc at the critical energy is bremsstrahlung or (energy-independent) diffusion (both of which favour an index of 0.6), with a sub-dominant secondary contribution from inverse Compton (which drags the index slightly upward). While this is the primary explanation for the spectral index, there are two additional factors that also contribute. First, while emission from the disc is dominant, emission from the halo is non-negligible, and we find that the spectral indices we obtain for halo emission are significantly softer (i.e., larger $\alpha_\mathrm{610\,MHz}^\mathrm{1.49\, GHz}$) than those from the disc. We show this in \autoref{fig:alpha_dischalo}. 
The halo spectral index is
softer
for two reasons: first, the effective CR injection index into the halo is much softer than that for primary and secondary injection in the disc due to greater losses suffered by higher-energy electrons while escaping the disc (\citetalias{roth+2023}), and, second, halo losses are completely dominated by inverse Compton emission. Thus the net effect of the halo contribution is also to push the spectral index upward away from the value of 0.6 favoured by bremsstrahlung or diffusion within the disc.

A final sub-dominant factor affecting the spectral index distribution is the effects of free-free opacity. While these are omitted from \autoref{fig:alpha_lossbins} for clarity, our findings illustrated in \autoref{fig:alphaSFR} suggest that, within the 610 MHz to 1.49 GHz frequency range we are considering, the characteristic hardening of the spectrum due to free-free emission is mostly important towards lower star-formation rates where free-free outshines weak radio synchrotron emission. The effect of free-free is to create a small population of low $\alpha$ galaxies from what would otherwise be a very narrow range of spectral index at low star formation rate, where galaxies are completed dominated by diffusive losses into the halo.
By contrast, free-free is unimportant at high star formation rates at the frequencies shown. While free-free absorption can become important at high star formation rates, its effects are not expected to be apparent until frequencies well above a few GHz except in the most compact and opaque galaxies, cf. the results for Arp 220 in \citetalias{roth+2023}.

\subsection{A predicted correlation between radio spectral shape and proton calorimetry}
\label{ssec:fcal}

We have now established how the radio spectral index is produced by the superposition of thermal free-free and non-thermal radio synchrotron emission. 
At this stage it is relevant to remember that, given the pronounced peaking
of the synchrotron spectrum at the critical frequency,
for a fixed observing frequency, we are probing the cosmic ray electron spectrum at approximately a single energy.
Observed radio spectra often have intrinsic curvature as a result of the mixing of components with different spectral indices at different frequencies \citep[e.g.,][]{2018A&A...611A..55K}.
This effect, which
can be seen in individual model galaxies plotted in \citetalias{roth+2023}, 
is particularly pronounced in the transition to thermal free-free dominance towards higher frequencies.
However, radio curvature can also reflect an underlying curvature in the steady state spectrum of the synchrotron-emitting electrons.
Such curvature  results, in turn, from the transitions between different loss mechanisms, dominant in different energy regimes.

As an example, if we take the spectral range between 610 MHz and 1.49 GHz, 
our model suggests that,
for low SFR systems, we can expect to be dominated by diffusive losses in the disc at low frequencies and a stronger contribution from free-free emission at high frequencies, which eventually dominates well beyond 10s of GHz.
This transition leads to a substantial {\it hardening} in the radio spectrum.
At the same time, such diffusion dominated systems feature low ISM densities and thus low (proton and electron) calorimetry fractions. 
At the opposite, high end of the SFR distribution, galaxies feature a high density ISM with high calorimetry fraction and losses that are dominated by ionisation at low frequencies and bremsstrahlung at high frequencies.
Transition between these leads to a {\it softening} of the spectrum (cf., the loss time plots in \citetalias{roth+2023}).

\begin{figure}
	\includegraphics[width=\columnwidth]{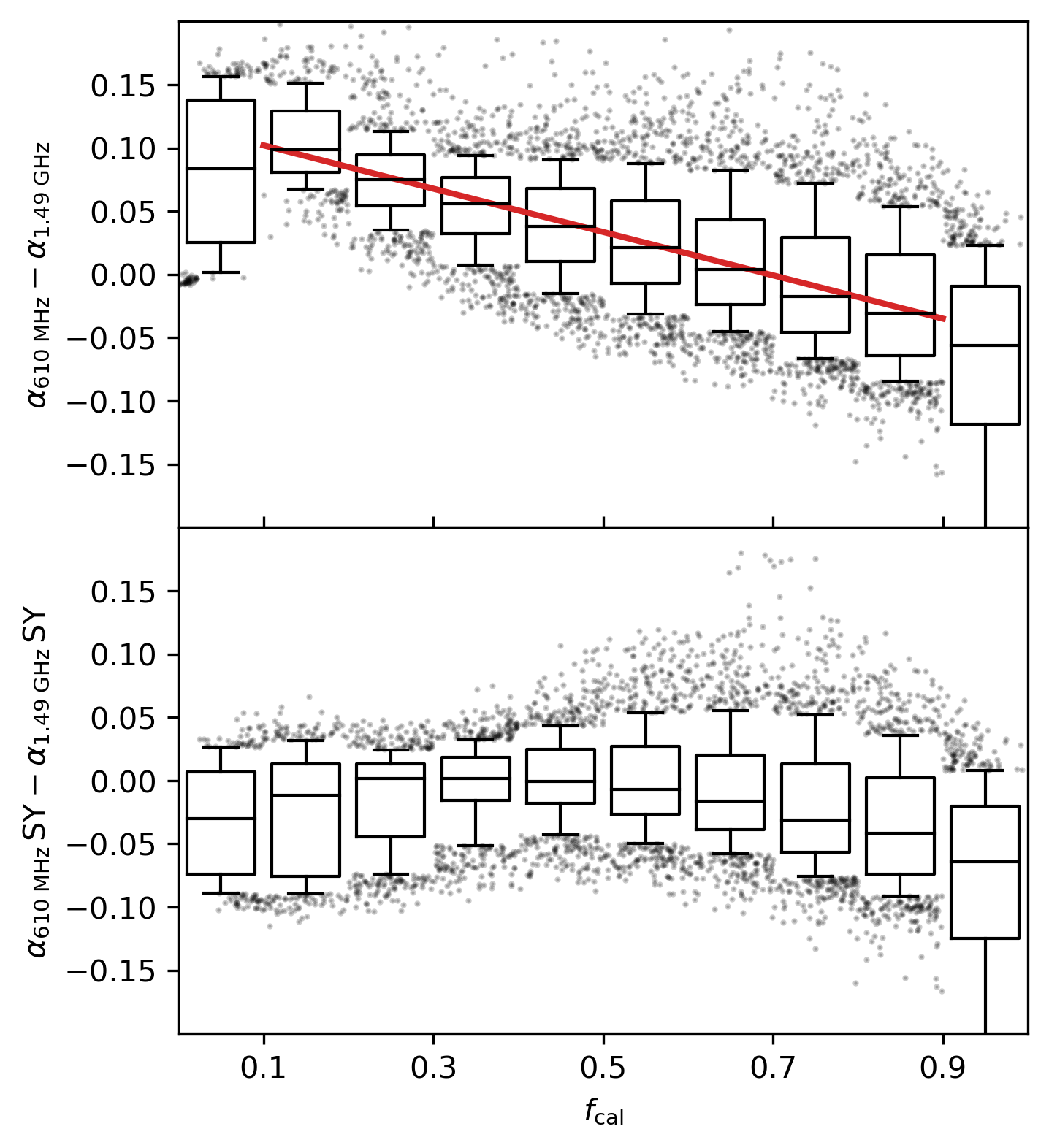}
    \caption{In the upper panel we show the change in spectral index between 610 MHz and 1.49 GHz, i.e.,$\Delta\alpha \equiv \alpha_{\rm 610 \; MHz} - \alpha_{\rm 1.49 \; GHz}$, as a function of proton calorimetry fraction $f_\mathrm{cal}$.
    Given our adopted convention $S_\nu \propto \nu^{-\alpha}$,
    a change $\Delta \alpha > 0$ corresponds to hardening, and $< 0$ corresponds to softening.
    The red line is a linear fit to the data in the indicated range.
    In the lower panel we show the same quantity but  considering only synchrotron emission (i.e., omitting free-free). }
    \label{fig:alphadelta2fcal}
\end{figure}

These considerations motivate us to investigate the correlation between the curvature of the radio spectrum, as parameterised by the change in spectral index from 610 MHz to 1.49 GHz $\Delta \alpha = \alpha_{\rm 610 \; MHz} - \alpha_{\rm 1.49 \; GHz}$, and the proton calorimetry fraction $f_{\rm cal}$, defined as the fraction of protons that produce pions within the ISM rather than escaping to the galactic halo (see Equation 17 of \citetalias{roth+2023} for a precise definition); we compute $f_{\rm cal}$ for protons  in the low energy limit where $f_{\rm cal}$ is essentially constant up to proton kinetic energies $\lesssim 10$ GeV.
%
We show the correlation between these quantities in the upper panel of \autoref{fig:alphadelta2fcal}, where  points $\Delta \alpha > 0$ correspond to spectral hardening with frequency (concave up spectrum), and points $\Delta \alpha < 0$ correspond to softening (concave down).
The resulting correlation is surprisingly tight; a linear fit gives $\Delta \alpha = -0.17 f_{\rm cal} + 0.12$, shown as a red line in the Figure. 
In the plot, the evident turn-down at the very low end of the calorimetry range corresponds to galaxies that are entirely dominated by thermal free-free emission in the spectral band considered and which therefore exhibit the corresponding fixed, hard spectral index characteristic of free-free, cf. \autoref{fig:alphaSFR}. This is not because free-free emission is exceptionally bright but rather because synchrotron emission in these galaxies is exceptionally dim.
The significant softening of the spectral index when approaching full calorimetry is explained with
a transition from
ionisation losses (which produce a very hard spectrum) dominating at 610 MHz 
to bremsstrahlung and other processes that produce a softer spectral index at 1.49 GHz.
Again we note that free-free {\it absorption} also helps to harden the low frequency spectrum for the most compact systems (cf. the `corrected' results for Arp 220 in \citetalias{roth+2023}). 

In the lower panel of \autoref{fig:alphadelta2fcal} we show the same data with the free-free emission component removed. It is evident that, while we recover the same spectral index change at high calorimetry fractions, the 
low $f_{\rm cal}$ behaviour is different.
This indicates that, for the overall radio spectrum, the total spectral index change is  controlled by the thermal free-free contribution for intermediate and low $f_{\rm cal}$ systems.
%


The results shown in the upper panel of \autoref{fig:alphadelta2fcal} represent a testable prediction of our model, which is one of the first to make population-level predictions for galactic $\gamma$-ray and radio spectra within a single unified framework. Actually performing the test, however, will require future instrumentation. Measurements of the radio spectral index at 610 MHz and 1.49 GHz are relatively straightforward to obtain for large samples, but estimates of the calorimetry fraction are more challenging, since they require measuring the $\gamma$-ray spectrum with enough fidelity to estimate the total power radiated via pion decay, and then dividing this by the star formation rate. As noted in \autoref{ssec:sampling}, at present estimates of the calorimetry fraction are available for only ten galaxies, with large uncertainties, and several of these have unusual radio behaviour (e.g., M31's likely pulsar-driven excess) that may well alter their spectral shapes compared to those of main sequence galaxies. However, with the upcoming much more sensitive \textit{Cherenkov Telescope Array} \citep{CTA19a}, the sample of $\gamma$-ray detected star-forming galaxies should expand considerably, and it should become possible to search for the correlation predicted in the upper panel of \autoref{fig:alphadelta2fcal}.

\subsection{Uncertainties and model caveats}

Due to the simplicity of our model, there are a number of uncertainties that should be pointed out. One is that we use a single radius -- the half-light radius (effective radius) at 5000 \AA\ -- for the radial dimension of the galactic disc. This choice is motivated by practical considerations, in that it is the only radius available for very large galaxy samples that are homogeneously-selected to cover the SFMS. In reality, of course, conditions vary radially within galaxies, and even if we limit ourselves to considering a single galactic radius there are a number of possible definitions \citep[e.g.,][]{Trujillo20a}. It is therefore worth considering the impact of alternate choices, since the star formation rate per unit area in our models, and therefore the radiation energy density and thence the inverse Compton loss time, scale as the inverse of radius squared. To this end, we first note that in local galaxies the star formation surface density is observed to be roughly exponentially distributed in galactocentric radius with a scale length $\approx 0.5 R_{25}$ \citep{Bigiel12a}, where $R_{25}$ is the usual radius of the 25 mag arcsec$^{-2}$ isophote in B-band.\footnote{Formally \citet{Bigiel12a} measure the scale length of molecular gas. However, since the star formation rate per unit molecular gas mass is observed to be nearly constant in the same sample \citep{Leroy13a}, the distribution of star formation is the same.} The relationship between $R_{25}$ and $R_{\rm e}$, the radius we take from CANDELS, varies with galaxy sample, but typically $R_{\rm e}$ is s smaller by a factor of a few \citep{Elmegreen98a}, which suggest that the radius we have chosen should on average be close to the scale radius of star formation. This implies that our estimate of the IC loss time is probably off by at most a factor of $\sim 2$. Even if this is the case, though, the effects on the FRC are likely to be small, simply because we find that IC losses in the disc are almost always subdominant for the relatively low-energy electrons whose synchrotron critical frequency is close to 1.49 GHz. Thus a factor of $\approx 2$ error in the IC loss rate would not materially alter our results. However, this is not to say that there cannot be exceptions in galaxies with very unusual properties. We discuss on such exception in \citetalias{roth+2023}: Arp 220, a major merger with a star formation radius that is only $\approx 10\%$ of its optical radius, a discrepancy so large that by using the optical radius we underestimate the free-free optical depth of the galaxy. This is very much the exception rather than the rule, however, since major mergers like Arp 220 represent a small contribution to the total star formation budget of the Universe at all redshifts \citep{Rodighiero11a}.

A second limitation is that our model currently does not treat ionisation losses for CR protons. This matters very little for the $\gamma$-ray emission produced by CRp because this emission comes almost exclusively from protons above the pion production threshold, and ionisation losses are negligible at such high energies. However, if we were interested in the shape of the proton CR spectrum at lower energies, ionisation losses would need to be treated in a similar vein to our treatment for CRe.

We finally remind the reader that, for simplicity, our model does not explicitly address non-thermal emission by heavy ions in the primary CR beam.

\section{Conclusion}
\label{sec:conclusion}

In this second paper of the \textsc{CONGRuENTS} series we have demonstrated that a model that relies on a few  structural parameters of galaxies -- viz., the stellar mass, the star-formation rate, and the half-light radius  -- can be used successfully to predict the non-thermal emission properties of galaxy populations. We show that applying the \textsc{CONGRuENTS} code to a realistic population of galaxies drawn from the CANDELS survey allows us to reproduce the observed far infrared-radio correlation (FRC), the observed far infrared-$\gamma$-ray correlation (F$\gamma$C), and the observed distribution of radio spectral indices.
We also make predictions for additional scaling relationships that have not yet been probed accurately
due to the limitations of current $\gamma$-ray instrumentation.
In particular, we anticipate refinements to the F$\gamma$C and the radio-$\gamma$ relations, and we show that a strong correlation should exist between the curvature of a galaxy's radio spectrum and the degree of proton calorimetry it achieves, as characterised by its $\gamma$-ray luminosity per unit star formation.
Our model achieves these things by integrating elements of a number of prior theoretical efforts, but extending their explanatory power by carefully modelling injection and loss mechanisms on a galaxy-by-galaxy basis. This allows us to gain deeper insights than previously possible. Theoretical efforts so far have included assumptions of nearly complete CR electron calorimetry with synchrotron emission as the dominant loss channel (e.g.,~\citealt{1989A&A...218...67V}) and models that assume near equipartition between synchrotron and inverse Compton losses (e.g.,~\citealt{2013A&A...556A.142S}) due to the nature of the turbulent dynamo \citep{1996A&A...306..677L, 2006ApJ...645..186T, 2010ApJ...717....1L}. These models have their limitations in that they do not predict the correct radio spectral indices and fail to predict effects due to ionisation and bremsstrahlung losses (see \autoref{sec:introduction}).
We recover full calorimetry,
in a broad sense,
for CR electrons for all galaxies in our sample, but in contrast to classical calorimeter theory our results suggest that synchrotron losses are always subdominant; a larger fraction of electron power is lost into other channels, particularly inverse Compton emission in the halos of low SFR systems and bremsstrahlung in the  discs of high SFR systems. 
The reason that the FRC is a relatively tight, featureless powerlaw despite galaxies not being {\it synchrotron} calorimeters is primarily that both the fraction of emission contributed by secondary electrons and the fraction of power radiated into synchrotron emission evolve only very weakly with star formation rate. Weak evolution of the synchrotron emission fraction is due to a smooth changeover between domination by diffusive losses into the halo (and thence into inverse Compton emission) at low star formation rates to domination by bremsstrahlung at higher star formation rates.

While a cancellation of this type has been proposed before, our results show that it is not simply an accident or a conspiracy, but is instead the inevitable result of basic physical principles and galaxy scaling relations. In particular, we show that hydrostatic balance and the turbulent dynamo combine to place tight constraints on galaxies' densities and magnetic field strengths, and that these -- plus the empirical scalings between galactic star formation rates, stellar masses, sizes, and gas velocity dispersions -- are sufficient to explain the FRC. The fact that the model simultaneously reproduces the distribution of galactic radio spectral indices provides an additional cross-check on its accuracy.
The success of the model is, in part, due to the detailed accounting it provides for the full range of CR electron energy loss processes, which also allows for decomposition between the various processes that end up shaping the steady state cosmic ray distributions and their resulting, non-thermal emission. This decomposition is also the basis for the new correlation we predict between proton calorimetry and radio spectral curvature.
One important implication of our work is that, when it comes to population-level correlations between thermal and non-thermal emission like the FRC, sample selection really matters. Because these correlations are in part shaped by empirical trends in galaxy populations -- most notably the existence of a star-forming main sequence (SFMS) and the Kennicutt-Schmidt relation and its extensions -- we do not necessarily expect the correlations to look exactly the same for all possible galaxy samples. For example, samples that preferentially select galaxies that fall away from the SFMS may show different correlations than samples that target more typical galaxies on it.
Our model makes clear, testable predictions for additional trends in the non-thermal emission of galaxy populations, and observations of these have the potential to break degeneracies between our proposal and existing models, but observational searches for these trends must be underpinned by the availability of observational data with well-understood sampling functions. We urge observers to carefully consider the selection of their sample and survey design so that the data can be compared to models in an unbiased manner.

Understanding sample selection and bias, and how these influence correlations between thermal and non-thermal emission, will be a key focus of a follow-up paper in which we intend to address the redshift evolution of the FRC. Selection effects are particularly likely to be important for this trend since, for instance, observations even of typical galaxies will tend to prove higher SFRs at higher redshifts due to evolution of the SFMS. This, in turn, means that observations at fixed (emitted-frame) frequency will probe lower cosmic ray energies. At the same time, for observations at a fixed (observed-frame) frequency, the critical energy increases as redshift increases. These two contrasting trends are likely to play an important role in determing FRC redshift evolution. Similarly, the evolution of the SFMS and galaxy gas richness with redshift will have profound implications for the balance between magnetic and radiation energy densities, and between the FIR and CMB components of the radiation energy density.

Beyond evolution of the FRC, there are a number of additional problems of interest and potentially useful applications and extensions for \textsc{CONGRuENTS}. The code currently does not include a contribution from non-thermal emission from millisecond pulsars, which has been shown to be a likely source of considerable radio synchrotron and $\gamma$-ray emission in low redshift, quenched galaxies \citep{2021PhRvD.103h3017S,2022NatAs...6.1317C}. This may have an interesting effect on radio-dependent correlations at the very lowest star-formation rates, or for galaxies well below the SFMS.

Another avenue we intend to explore is to extend the results obtained in \citet{2021Natur.597..341R} for the diffuse, isotropic $\gamma$-ray background to derive the contribution of star-forming galaxies to the diffuse backgrounds of neutrinos and radio \citep[e.g.,][]{2021APh...12602532N}, and to predict correlations between galactic neutrino emission and radio and $\gamma$-ray emission, as we have predicted radio-$\gamma$-ray relationships here.
There are a number of instruments in various stages of realisation, such as CTA, SKA and IceCube Gen 2, which will significantly increase sensitivity in all three of these bands and messengers in the near future. It is very much expected that these instruments will provide the ultimate test for our predictions in this work.

\section*{Acknowledgements}
The authors acknowledge support from the Australian Research Council through its \textit{Discovery Projects} funding scheme, award DP230101055. MAR acknowledges support from the Australian National University through a research scholarship (MAR). This research was undertaken with the assistance of resources and services from the National Computational Infrastructure (NCI), which is supported by the Australian Government. TAT was partially supported by NASA grant\,80NSSC18K0526.
\section*{Data Availability}

The data underlying this article are available in Zenodo, at https://doi.org/10.5281/zenodo.10892689. Where external data were used, these are publicly accessible and are referenced in the text.

\section*{Code Availability}

\textsc{CONGRuENTS}, the code used to compute the results of this research is open source and available at https://doi.org/10.5281/zenodo.7935188.



\bibliographystyle{mnras}
\bibliography{bibliography} 




\begin{appendix}


\end{appendix}

\bsp	
\label{lastpage}
\end{document}